\documentclass{article}

\usepackage{arxiv}

\usepackage[utf8]{inputenc}
\usepackage[T1]{fontenc}
\usepackage{hyperref}
\usepackage{url}
\usepackage{booktabs}
\usepackage{amsmath,amssymb}
\usepackage{graphicx}
\graphicspath{{./}}
\usepackage{natbib}
\usepackage{array}
\usepackage{float}
\usepackage{longtable}
\usepackage{multirow}
\usepackage{subcaption}
\usepackage{textcomp}
\usepackage{calc}
\usepackage{tikz}
\usetikzlibrary{arrows.meta,backgrounds,calc,fit,positioning,shapes.geometric}
\usepackage{xcolor}

\definecolor{phase1fill}{HTML}{DBEAFE}
\definecolor{phase1draw}{HTML}{3B82F6}
\definecolor{phase2fill}{HTML}{D1FAE5}
\definecolor{phase2draw}{HTML}{10B981}
\definecolor{phase3fill}{HTML}{FEF3C7}
\definecolor{phase3draw}{HTML}{F59E0B}
\definecolor{phase4fill}{HTML}{EDE9FE}
\definecolor{phase4draw}{HTML}{8B5CF6}
\definecolor{clientfill}{HTML}{DBEAFE}
\definecolor{clientdraw}{HTML}{3B82F6}
\definecolor{backendfill}{HTML}{FEF3C7}
\definecolor{backenddraw}{HTML}{F59E0B}
\definecolor{apifill}{HTML}{EDE9FE}
\definecolor{apidraw}{HTML}{8B5CF6}
\definecolor{datafill}{HTML}{D1FAE5}
\definecolor{datadraw}{HTML}{10B981}

\makeatletter
\newsavebox\pandoc@box
\newcommand*\pandocbounded[1]{%
  \sbox\pandoc@box{#1}%
  \Gscale@div\@tempa{\textheight}{\dimexpr\ht\pandoc@box+\dp\pandoc@box\relax}%
  \Gscale@div\@tempb{\linewidth}{\wd\pandoc@box}%
  \ifdim\@tempb\p@<\@tempa\p@
    \let\@tempa\@tempb
  \fi
  \ifdim\@tempa\p@<\p@
    \scalebox{\@tempa}{\usebox\pandoc@box}%
  \else
    \usebox{\pandoc@box}%
  \fi
}
\makeatother

\makeatletter
\@ifpackageloaded{caption}{}{\usepackage{caption}}
\AtBeginDocument{%
\ifdefined\contentsname
  \renewcommand*\contentsname{Table of contents}
\else
  \newcommand\contentsname{Table of contents}
\fi
\ifdefined\listfigurename
  \renewcommand*\listfigurename{List of Figures}
\else
  \newcommand\listfigurename{List of Figures}
\fi
\ifdefined\listtablename
  \renewcommand*\listtablename{List of Tables}
\else
  \newcommand\listtablename{List of Tables}
\fi
\ifdefined\figurename
  \renewcommand*\figurename{Figure}
\else
  \newcommand\figurename{Figure}
\fi
\ifdefined\tablename
  \renewcommand*\tablename{Table}
\else
  \newcommand\tablename{Table}
\fi
}
\@ifpackageloaded{float}{}{\usepackage{float}}
\floatstyle{ruled}
\@ifundefined{c@chapter}{\newfloat{codelisting}{h}{lop}}{\newfloat{codelisting}{h}{lop}[chapter]}
\floatname{codelisting}{Listing}

\makeatother
\makeatletter
\@ifpackageloaded{caption}{}{\usepackage{caption}}
\@ifpackageloaded{subcaption}{}{\usepackage{subcaption}}
\makeatother

\title{Can AI Help You Get Over Your Breakup? \\
One Session with a Belief-Reframing Chatbot Shows Sustained Distress Reduction}

\author{
  Thomas Menzel\thanks{Authors contributed equally to this work and are listed alphabetically. Corresponding authors: \texttt{thomaswmenzel@gmail.com} (TM), \texttt{ms2957@cam.ac.uk} (MS).} \\
  Technical University of Munich, Germany \\
  University of Cambridge, United Kingdom \\
  \And
  Michel Schimpf\footnotemark[1] \\
  University of Cambridge, United Kingdom \\
   \And
  Thomas Bohné \\
  University of Cambridge, United Kingdom \\
}

\renewcommand{\shorttitle}{AI Chatbot for Breakup Distress}

\hypersetup{
pdftitle={Can AI Help You Get Over Your Breakup? One Session with a
Belief-Reframing Chatbot Shows Sustained Distress Reduction},
pdfauthor={Thomas Menzel, Michel Schimpf, Thomas Bohné},
pdfkeywords={breakup distress, cognitive reappraisal, memory
reconsolidation, AI chatbot, randomized controlled trial, digital mental
health},
}

\begin{document}
\maketitle

\begin{abstract}
Romantic breakups are among the most common and intense sources of
psychological distress. We evaluated \emph{overit}, a single-session AI
chatbot that uses cognitive reappraisal to address breakup distress,
informed by memory reconsolidation theory. In a
pre-registered\footnote{Pre-registered at
  \url{https://doi.org/10.17605/OSF.IO/43MJ9}} randomized controlled
trial, 254 adults in the United States and United Kingdom who had
experienced a romantic breakup were assigned to either an initial survey
assessment followed by an AI chat session or to a survey-only control.
Breakup distress was measured at baseline, 7 days, and again at an
exploratory 1-month follow-up using the Breakup Distress Scale.
Participants assigned to \emph{overit} showed a significantly greater
reduction in breakup distress than controls at 7 days (time-by-condition
interaction \(B = -5.36\), \(SE = 1.19\), \(p < .001\); completer-based
\(d = -0.70\)). A smaller but still significant treatment advantage
remained detectable at the exploratory 1-month follow-up among
post-session completers (\(B = -2.92\), \(SE = 1.22\), \(p = .017\)).
Exploratory post hoc moderation suggested a larger effect among male
participants (\(B = 7.78\), \(p = .003\)). These results suggest that a
brief AI chatbot conversation can meaningfully reduce breakup distress,
with exploratory evidence that a smaller advantage persists over the
following month. Future work should test the intervention against active
controls, evaluate repeated-session use, and recruit more diverse
samples.
\end{abstract}


\section{Introduction}\label{sec-introduction}

Romantic breakups are common and often cause psychological distress.
Most adults report at least one breakup across the life course, and
breakup distress is associated with depressive symptoms, anxiety,
intrusive thoughts, and sleep disruption
\citep{perillouxBreakingRomanticRelationships2008, fieldBreakupDistressUniversity2009, verhallenRomanticRelationshipBreakup2019}.

One plausible intervention target is breakup-related cognitions, which
may contribute to maintaining distress beyond healthy grief periods
\citep{boelenCognitiveBehavioralConceptualization2006}. Negative beliefs
about the self, the former relationship, and one's own emotional
reactions are associated with breakup-related grief, depression, and
distress
\citep{boelenNegativeCognitionsEmotional2009, slotterWhoAmYou2010, brennerMeasuringThoughtContent2015}.
Cognitive reappraisal is a method that addresses this level of meaning by
helping individuals reinterpret emotionally salient events, and
breakup-specific experiments suggest that reframing the ex-partner or
the relationship can reduce attachment-related distress
\citep{langeslagDownregulationLoveFeelings2018}. More broadly, cognitive
reappraisal and cognitive restructuring are linked to resilience and
therapeutic improvement across clinical contexts, including digital
interventions
\citep{stoverMetaanalysisCognitiveReappraisal2024, ezawaCognitiveRestructuringPsychotherapy2023, morelloCognitiveReappraisalMHealth2023}.
In practice, cognitive reappraisal asks people to surface a distressing
interpretation, weigh the evidence for it, and arrive at one that fits
the situation better
\citep{beckCognitiveTherapyDepression1979, ezawaCognitiveRestructuringPsychotherapy2023}.
This work is uncomfortable by design, because changing a meaning means
challenging it rather than agreeing with it.

Memory reconsolidation provides one theoretical account of how such
revision might produce durable change. Memory reconsolidation theory
proposes that when an emotional learning is reactivated, it becomes
temporarily labile and can be updated if the person enters an emotional
state that contradicts what the original learning predicts
\citep{naderReconsolidationDynamicNature2015, eckerUnlockingEmotionalBrain2012, sevensterPredictionErrorGoverns2013}.
Clinical reconsolidation accounts therefore emphasize a sequence in
which a painful schema is first reactivated, then juxtaposed with
discrepant meaning, and finally experienced as less compelling or less
emotionally dominant
\citep{laneMemoryReconsolidationEmotional2015, eckerMemoryReconsolidationUnderstood2015}.
This framework is especially relevant for breakup distress because many
breakup-related beliefs are narrative and relational
\citep{neimeyerReconstructingMeaningBereavement2011, slotterWhoAmYou2010}.
A person may hold beliefs such as ``I was abandoned because I am not
enough'' or ``this breakup proves I will not recover''
\citep{boelenNegativeCognitionsEmotional2009}. A structured conversation
can guide users through articulating those beliefs, examining them in
context, generating conflicting interpretations, and consolidating any
emerging shift in perspective. Brief reconsolidation-informed
reinterpretation paradigms have shown sustained changes in emotional
memory, suggesting that even a single-session intervention may be
capable of producing clinically meaningful change
\citep{speerFindingPositiveMeaning2021, eckerMemoryReconsolidationUnderstood2015}.

Chatbots are a plausible delivery format for this kind of intervention
because they can offer private, conversational guidance in the moments
when distress is highest. People already turn to large language models
for emotional support, and such systems are often experienced as
empathic and responsive
\citep{mcbainUseGenerativeAI2025, ayersComparingPhysicianArtificial2023, sorinLargeLanguageModels2024}.
But an empathic tone is not enough for therapeutic benefit. Modern LLMs
are trained in ways that bias them toward agreement and over-validation,
a tendency documented as sycophancy
\citep{ouyangTrainingLanguageModels2022, perezDiscoveringLanguageModel2023, sharmaTowardsUnderstandingSycophancy2024, malmqvistSycophancyLargeLanguage2024}.
That is a problem when distressed users present rigid or catastrophic
interpretations that need to be challenged rather than affirmed, and
recent evaluations of AI therapy bots show that unstructured systems can
respond unsafely or reinforce problematic beliefs in clinical scenarios
\citep{mooreExpressingStigmaInappropriate2025, auYeungPsychogenicMachineSimulating2025}.
A useful chatbot-delivered reappraisal intervention for breakup distress
therefore cannot just listen and validate; it has to help users revisit
and revise the meanings that keep distress active. No study has yet
tested whether a structured, reconsolidation-informed AI chatbot can do
this in a randomized controlled trial.

In the present study, we tested whether \emph{overit}, an LLM-based
belief-reframing chatbot for breakup support, could reduce breakup
distress at 7-day follow-up relative to an assessment-only control
condition (primary, confirmatory), and whether post-session insight
mediated change, consistent with the hypothesized emotional juxtaposition
that reconsolidation-informed reappraisal posits (exploratory). We also
exploratorily analyzed whether the treatment advantage remains stable at
an (on average) 36-day follow-up, and whether intervention effects varied by
baseline characteristics, including attachment style measures
\citep{gehlAttachmentBreakupDistress2024, eckerMemoryReconsolidationUnderstood2015, sevensterPredictionErrorGoverns2013}.
The broader question is whether a single structured LLM-based chatbot
conversation can function as an effective and potentially durable
intervention for breakup distress and whether reconsolidation-informed
approaches can work for psychological variables beyond trauma
research, where it has mostly been applied.

\section{Methods}\label{sec-methods}

\subsection{Participants and Design}\label{participants-and-design}

To test whether a single AI chatbot session could reduce breakup
distress, we conducted a preregistered, two-arm randomized controlled
trial. Participants were assigned to either a brief chatbot intervention
or an assessment-only control condition, with breakup distress at 7-day
follow-up as the primary outcome.

Adults aged 18 years or older living in the United States or the United
Kingdom who had experienced the end of a romantic relationship were
recruited through Prolific. An a priori power analysis targeted a
small-to-moderate between-group difference in change on the Breakup
Distress Scale (BDS; \(d = 0.30\), two-sided \(\alpha = .05\), 80\%
power); allowing for attrition, the recruitment target was set to
\(N = 220\). An additional 1-month questionnaire was later distributed
to participants who had completed the day-0 post-session survey
(intervention participants who had finished both the chatbot conversation
and the post-chat survey, and control participants who had finished the
matched day-0 feedback survey) and was analyzed exploratorily.

\subsection{Intervention and Control
Conditions}\label{intervention-and-control-conditions}

\begin{figure}[htbp]

\begin{minipage}{0.33\linewidth}

\centering{

\pandocbounded{\includegraphics[keepaspectratio]{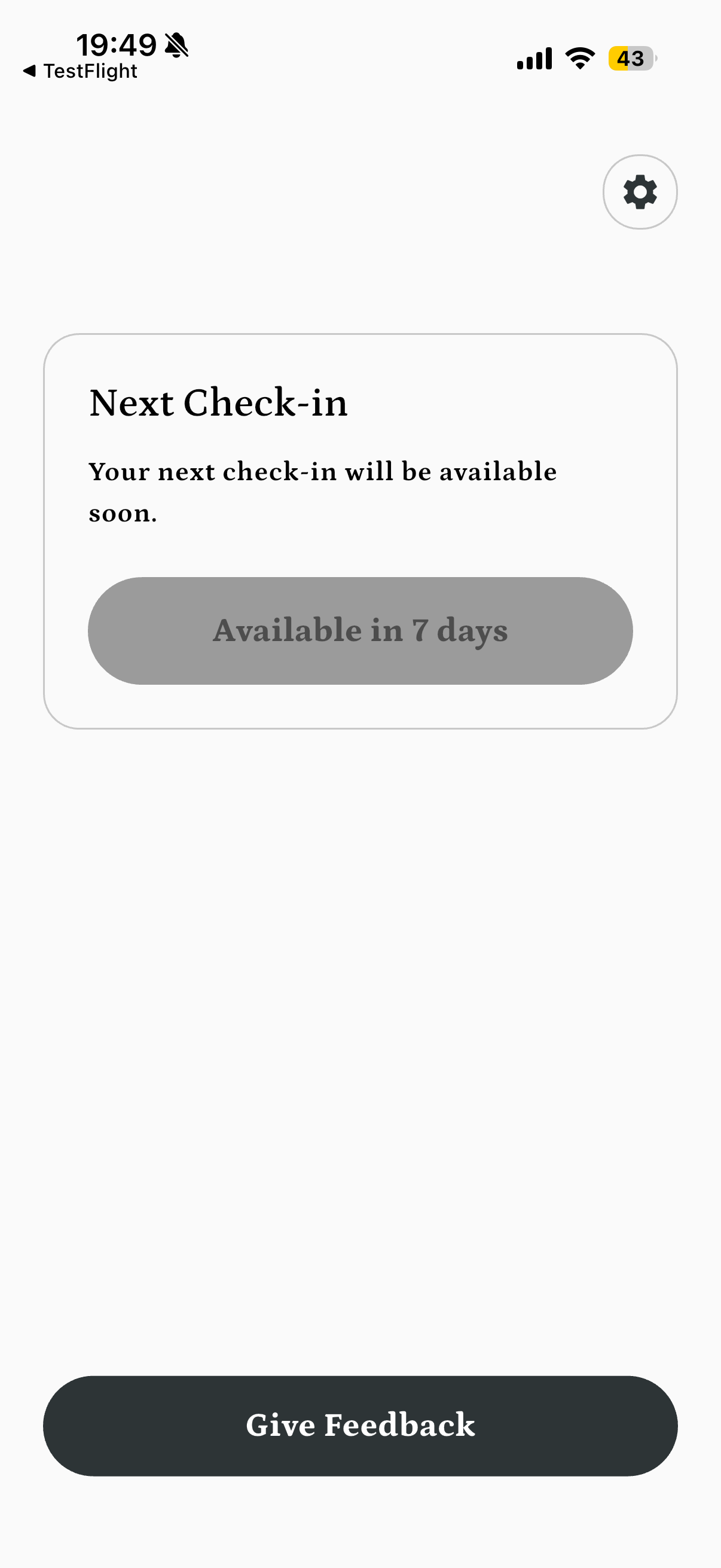}}

}

\subcaption{\label{fig-control-home}Control home screen}

\end{minipage}%
\begin{minipage}{0.33\linewidth}

\centering{

\pandocbounded{\includegraphics[keepaspectratio]{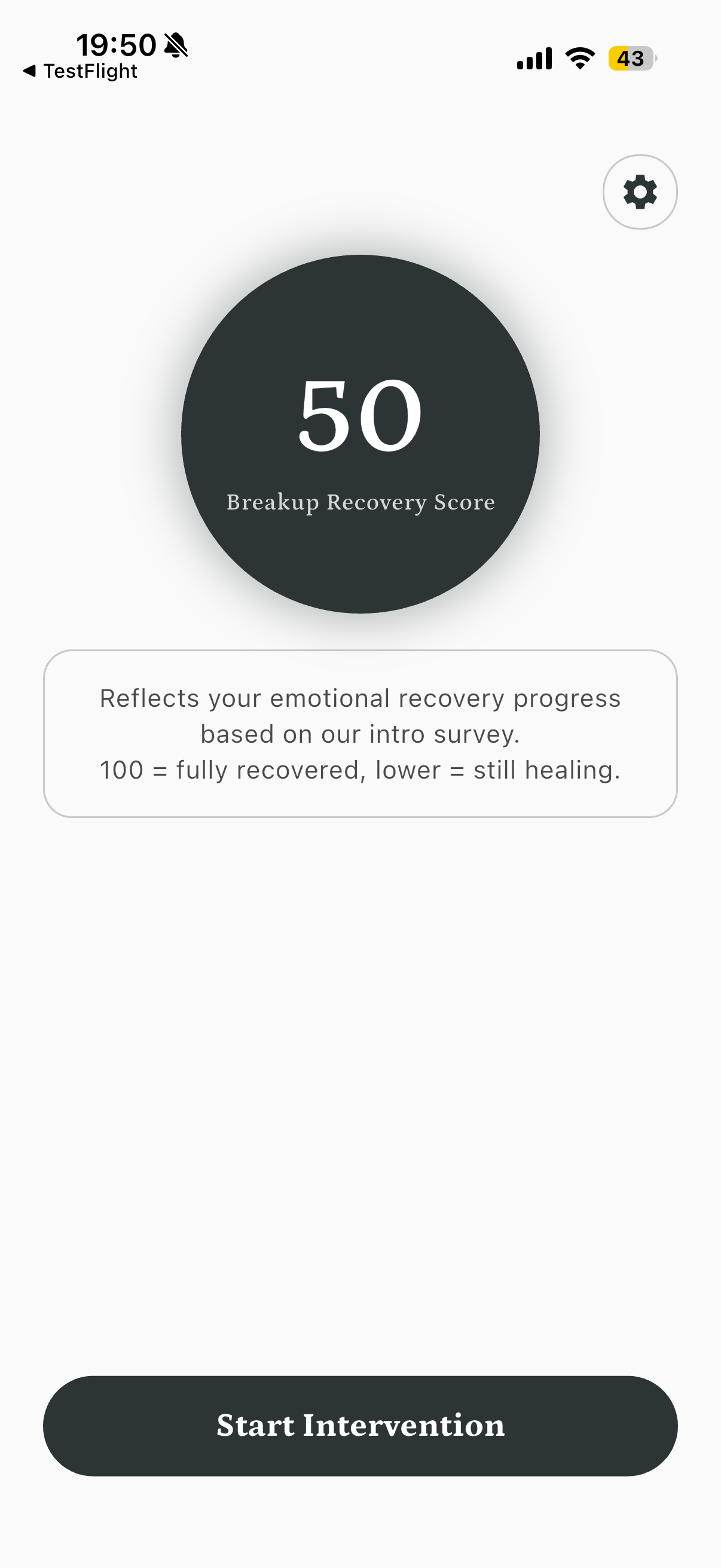}}

}

\subcaption{\label{fig-treatment-home}Treatment home screen}

\end{minipage}%
\begin{minipage}{0.33\linewidth}

\centering{

\pandocbounded{\includegraphics[keepaspectratio]{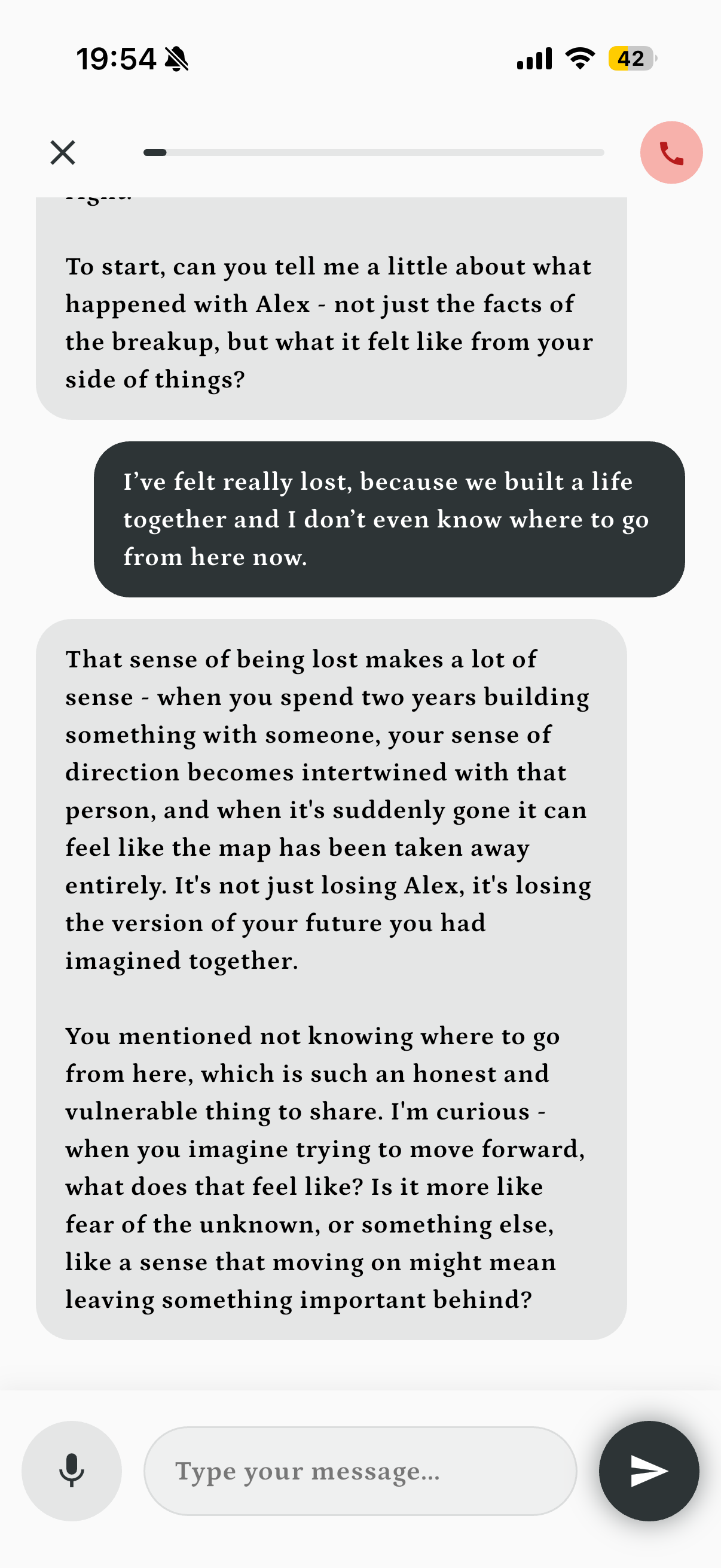}}

}

\subcaption{\label{fig-treatment-chat}Chat intervention}

\end{minipage}%

\caption{\label{fig-app-screenshots}App screenshots showing the control
group home screen, the treatment group home screen with breakup recovery
score, and the chatbot conversation interface.}

\end{figure}

Intervention participants used \emph{overit}, a single-session AI
chatbot for breakup recovery delivered through a mobile app. The chatbot
guided users through a structured conversation intended to reactivate
breakup-related memories and beliefs, examine limiting interpretations,
generate alternative perspectives, and integrate a revised understanding
of the experience. Conversations typically lasted about 20 minutes and
could be completed through text or voice input (Figure~\ref{fig-app-screenshots}).

The conversation followed four phases: context gathering, belief
exploration, counterfactual generation, and integration and closure.
These phases incorporated both insights from the Therapeutic Reconsolidation Process \citep{eckerUnlockingEmotionalBrain2012, eckerMemoryReconsolidationUnderstood2015}, as well as from classic cognitive restructuring \citep{beckCognitiveTherapyDepression1979, hofmannEfficacyCognitiveBehavioral2012}. They were designed to map onto a
sequence in which breakup-related meanings were first reactivated, then
examined and challenged, before participants were guided toward
alternative interpretations and a final summary of what had shifted
during the session. Table~\ref{tbl-phase-prompts} shows excerpts of the phase-specific
instructions appended to the base system prompt at each turn. The
wording changed with the current phase and milestone state.

\begin{longtable}[]{@{}
  >{\raggedright\arraybackslash}p{(\linewidth - 2\tabcolsep) * \real{0.1667}}
  >{\raggedright\arraybackslash}p{(\linewidth - 2\tabcolsep) * \real{0.8333}}@{}}
\caption{Excerpts of phase-specific instructions appended to the base system prompt
at each turn, alongside the current turn count, milestone status, and
the participant's baseline assessment
data.}\label{tbl-phase-prompts}\tabularnewline
\toprule\noalign{}
\begin{minipage}[b]{\linewidth}\raggedright
Phase
\end{minipage} & \begin{minipage}[b]{\linewidth}\raggedright
Example system prompt instruction
\end{minipage} \\
\midrule\noalign{}
\endfirsthead
\toprule\noalign{}
\begin{minipage}[b]{\linewidth}\raggedright
Phase
\end{minipage} & \begin{minipage}[b]{\linewidth}\raggedright
Example system prompt instruction
\end{minipage} \\
\midrule\noalign{}
\endhead
\bottomrule\noalign{}
\endlastfoot
1: Context gathering & ``Focus on understanding their breakup story and
emotional response. Listen for limiting beliefs they express about
themselves or the relationship. Ask open-ended questions about what
happened and how they're feeling. Pay attention to patterns in their
thinking (e.g., `I'm not good enough,' `I'll never find love
again').'' \\
2: Belief exploration & ``You MUST identify at least one core limiting
belief before moving forward. Ask direct questions about their
self-perception and conclusions about the breakup. Listen for absolute
statements, catastrophizing, or self-blame. Once identified, begin
exploring the evidence for and against this belief.'' \\
3: Counterfactual generation & ``You MUST help them generate at least
one counterfactual scenario. Introduce `What if\ldots{}' questions based
on their beliefs. Explore alternative explanations for events they've
interpreted negatively. Make counterfactuals feel safe and exploratory,
not dismissive of their pain.'' \\
4: Integration and closure & ``Help them articulate what they've learned
or realized today. Ask what feels different or new in their thinking.
Acknowledge their courage in exploring difficult emotions. Once
therapeutic work feels complete, give a warm, definitive closure
message.'' \\
\end{longtable}

The chatbot was personalized using baseline assessments and breakup
context, including BDS responses, breakup timing, and the former
partner's first name. Control participants did not interact with the
chatbot and instead completed the assessment flow and a matched
post-session survey only. Detailed system prompts and milestone-tracking
logic will be made available with the published version of this work.

\subsection{System Architecture}\label{system-architecture}

A single system prompt encoding an entire therapeutic protocol creates
two practical problems for an LLM-delivered intervention. First, LLM
instruction-following degrades as the number of simultaneous constraints
in the prompt grows
\citep{jiangFollowBenchMultilevelFinegrained2024}, particularly when
relevant instructions are not positioned at the start or end of the
input \citep{liuLostMiddleHow2024}. Second, the same RLHF
training that makes modern LLMs feel empathic also biases them toward
agreement, so a model asked to both validate the user and challenge
distorted beliefs from the same prompt tends to default to validation
\citep{sharmaTowardsUnderstandingSycophancy2024}. \emph{overit} therefore
follows a script-based dialog policy in the spirit of
\citet{wasenmullerScriptBasedDialog2024} and the LLM-as-judge paradigm
\citep{zhengJudgingLLMasJudge2023}, separating the model call
that produces the user-facing reply from the model call that evaluates
therapeutic progress~\citep{schimpfAIAssistedGoalSettingImproves2026}.

Each conversational turn issues two calls to Claude Sonnet
(claude-sonnet-4-5-20250929) through the Anthropic API. A
\emph{generation call} (temperature 1.0) produces the user-facing
response, conditioned on a system prompt assembled at runtime from a
shared base prompt, the participant's baseline assessment data
(BDS responses, breakup timing, former partner's first name), the
conversation history, and the phase-specific instructions for the
current phase only (excerpts in Table~\ref{tbl-phase-prompts}). A
parallel \emph{evaluation call} (temperature 0) reviews the last three
exchanges against five binary milestones: (1) a core limiting belief has
been identified, (2) that belief has been challenged, (3) a counterfactual
interpretation has been genuinely considered, (4) the participant has
articulated a new insight or shift in perspective, and (5) natural
closure has been reached. Phase transitions are primarily turn-based
within a session capped at 18 turns, with a milestone-gated requirement
that the core limiting belief is identified before the conversation can
advance past Phase 2. Because the generation call sees only the
instructions relevant to the current phase, no single prompt must
juggle empathy generation, structural pacing, and clinical
progress-checking simultaneously.

The mobile app was built with Flutter and distributed to participants
through Apple's TestFlight beta testing program, with a Flask backend on
Google Cloud Run handling prompt assembly, phase management, and the
two-call sequence above. Voice input was supported via OpenAI Whisper
served through the Groq API, and Firestore stored conversation state
under anonymous user IDs.

\subsection{Measures}\label{measures}

Breakup distress was assessed at baseline and 7-day follow-up with the
16-item Breakup Distress Scale {[}BDS;
\citet{fieldBreakupDistressUniversity2009}{]}. The later exploratory
1-month questionnaire repeated the same BDS items. Adult attachment was
measured at baseline with the Experiences in Close Relationships
Scale-Short Form {[}ECR-S;
\citet{weiExperiencesCloseRelationship2007}{]}, yielding
attachment-anxiety and attachment-avoidance scores. Additional baseline
variables included time since breakup, relationship duration, breakup
initiator, and whether the participant remained in contact with their
former partner.

Insight was examined with dichotomous items asking whether
participants had experienced a sudden insight or ``aha moment'' during
the session and again during the week before follow-up. Participants
who reported having a sudden insight were invited to describe it in an open-ended response.
The exploratory 1-month questionnaire also asked whether participants
had experienced a sudden insight about their breakup in the weeks after the conversation
and invited optional free-text feedback. Immediately after the AI conversation (for the control group, after filling out the introduction survey), participants completed the UMUX-Lite
\citep{lewisUMUXLITEWhenTheres2013}, four single-item ratings of
empathy/support, trust, safety, and human-likeness, a single-item
recommendation rating (0--10), and three open-ended feedback questions.
An attention-check item was embedded in the BDS at all follow-up waves
and was excluded from BDS scoring. Full scale properties are retained in
Appendix~\ref{sec-methods-appendix}.

\subsection{Procedure}\label{procedure}

The study was approved by the University of Cambridge ethics board.
Participants accessed the study through Prolific and were directed to
download a research iOS app distributed through Apple's TestFlight beta
testing program. After providing informed consent, they were randomized
to their respective conditions and completed baseline breakup-context
questions together with the BDS and ECR-S.

Intervention participants viewed their BDS score result, rescaled to a
0-100 `breakup recovery score', in line with measurement-based care
principles \citep{lewisImplementingMeasurementBased2019}, and completed
the chatbot session before the post-session survey, whereas control
participants proceeded directly to a matched survey with similar questions about the app experience in general. Seven days later,
participants were invited through Prolific to complete the follow-up
assessment, which included the BDS and the follow-up insight items,
after which they received a completion code for compensation. After the
main trial window, participants who had completed the day-0 post-session
survey (intervention participants who had finished both the chatbot
conversation and the post-chat survey, and control participants who had
finished the matched day-0 feedback survey; \(n = 203\)) were
recontacted through Prolific for an additional exploratory questionnaire
approximately 1 month later (M = 36 days) that repeated the BDS and
collected insight and optional feedback items.

\subsection{Statistical Analysis}\label{statistical-analysis}

All statistical analyses reported in this version were conducted in
Python 3 using \texttt{pandas}, \texttt{SciPy}, and
\texttt{statsmodels}. The confirmatory primary analysis used a linear
mixed-effects model with BDS score as the outcome, fixed effects for
time (baseline versus 7-day follow-up), condition (intervention versus
control), and their interaction, and a random intercept for participant.
The primary test was the time-by-condition interaction, estimated with
maximum likelihood so that participants with incomplete follow-up data
could still contribute baseline observations. Completer-based
change-score comparisons are reported descriptively to aid
interpretation. Post-session user-experience, mediation, and later
follow-up analyses used the subset with completed post-session surveys.

Exploratory mediation analyses tested whether post-session insight or
follow-up insight mediated the treatment effect on BDS change. The
binary insight mediator was modeled with linear probability regression
to keep both paths on the same additive scale, and indirect effects were
estimated as the product of coefficients with 10,000 bootstrap resamples.
Exploratory moderator analyses regressed BDS change on condition, the
centered moderator, and the interaction term for attachment anxiety,
attachment avoidance, time since breakup, relationship duration, breakup
initiator, and contact with the former partner. The exploratory 1-month
extension used a mixed-effects model across all available baseline,
7-day, and 1-month observations and evaluated the treatment-by-1-month
interaction. User-experience ratings were compared between conditions
with Welch t-tests, and open-ended feedback is summarized qualitatively.

\section{Results}\label{sec-results}

\subsection{Sample}\label{sample}

Of the 254 primary-sample participants (121 treatment, 133 control), 171
completed the 7-day Breakup Distress Scale follow-up (79 treatment, 92
control), corresponding to an attrition rate of 32.7\% overall. Loss to
follow-up did not differ significantly between conditions (treatment:
34.7\%; control: 30.8\%; \(p = .599\)). The primary mixed-effects model
retained all 425 baseline and follow-up observations from the 254
baseline-valid participants.

An additional exploratory 1-month questionnaire was sent to 173 of
the 203 day-0 post-session-survey completers (94 intervention completers
and 109 control completers; 82 treatment and 91 control responded at
1 month). Response rates at this later wave did not differ by condition
(87.2\% of treatment participants versus 83.5\% of control participants;
\(p = .581\)).

The mean age of the participants was 36.4 years (SD = 10.8), 69.6\% were
female, 30.0\% were male, and 70.9\% identified as White. Across the
full primary sample, the mean time since breakup was 17.8 months (SD =
29.3), 42.1\% reported still being in contact with their former partner,
and 31.5\% reported being in a new relationship.

Groups were comparable at baseline on the primary outcome and
preregistered moderator variables. Baseline breakup distress did not
differ between conditions (treatment: M = 35.29, SD = 10.23; control: M
= 35.96, SD = 11.53; \(p = .623\)), and the groups were similarly
balanced on attachment anxiety (\(p = .518\)) and attachment avoidance
(\(p = .745\)). Time since breakup was numerically lower in the
treatment group (15.0 months) than in the control group (20.4 months),
but this difference did not reach significance (\(p = .134\)).

\subsection{Primary Outcome}\label{primary-outcome}

\begin{figure}[htbp]

\centering{

\includegraphics[width=0.92\linewidth,height=\textheight,keepaspectratio]{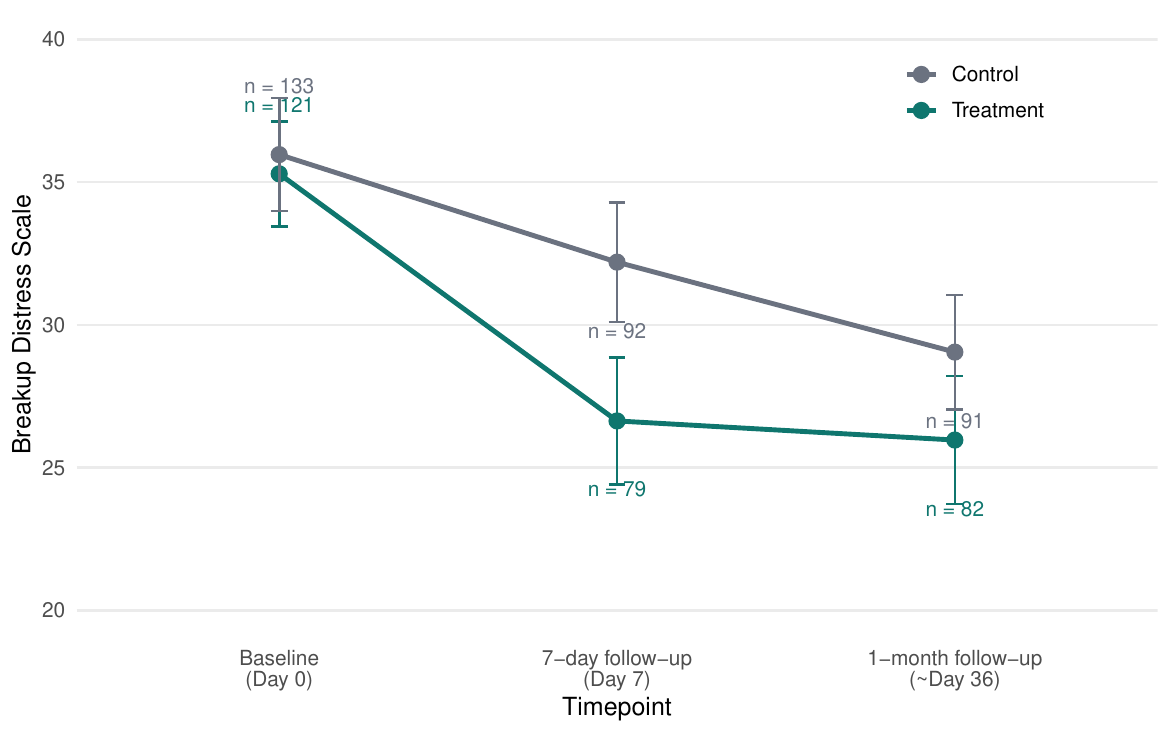}

}

\caption{\label{fig-primary-outcome}Breakup Distress Scale scores by
condition at baseline, the preregistered 7-day follow-up, and the later
exploratory 1-month follow-up. Baseline (treatment \emph{n} = 121,
control \emph{n} = 133) and 7-day (treatment \emph{n} = 79, control
\emph{n} = 92) values reflect the primary mixed-model sample. The
1-month values (treatment \emph{n} = 82, control \emph{n} = 91) reflect
observed responses to a later exploratory questionnaire, and the
baseline point is plotted from the full primary sample rather than from
the 1-month-responder subset. The 1-month point should therefore be
interpreted as a secondary follow-up extension rather than a third
confirmatory assessment wave. Error bars show 95\% confidence intervals
of the observed means.}

\end{figure}

The main analysis tested whether breakup distress changed differently
over time across conditions. A linear mixed-effects model with fixed
effects for time, condition, and their interaction, and a random
intercept for participant, showed a significant time-by-condition
interaction, \(B = -5.36\), \(SE = 1.19\), \(z = -4.50\), \(p < .001\),
95\% CI {[}-7.69, -3.03{]}. The treatment group therefore showed a
larger reduction in breakup distress from baseline to follow-up than the
assessment-only control group. Among 7-day responders, baseline scores
did not differ by condition (\(p = .991\)), and the treatment group
scored 5.56 points lower than controls at follow-up (26.63 vs.~32.20; Figure~\ref{fig-primary-outcome}).

Among the 171 participants who completed both assessments, the treatment
group showed a mean reduction of 9.23 points (SD = 9.34) on the Breakup
Distress Scale, compared with 3.68 points (SD = 6.58) in the control
group. The corresponding standardized effect size for completers was \(d = -0.70\). Any
reduction in breakup distress was observed in 84.8\% of treatment
participants and 68.5\% of control participants.

\subsection{Exploratory Results}\label{exploratory-results}

\subsubsection{1-Month Follow-up}\label{month-follow-up}

In an additional exploratory 1-month follow-up analysis, the treatment
advantage remained detectable but was smaller than at 7 days. A Python
mixed-effects model across all available baseline, 7-day, and 1-month
observations showed a significant 1-month treatment interaction,
\(B = -2.92\), \(SE = 1.22\), \(z = -2.39\), \(p = .017\), 95\% CI
{[}-5.31, -0.53{]}. Among the 173 1-month responders, the treatment
group showed a mean reduction of 9.84 points (SD = 10.71) on the BDS
from baseline, compared with 7.26 points (SD = 8.81) in the control
group, a difference in reductions of 2.58 points, \(d = -0.26\).
Observed mean BDS scores at 1 month were 25.96 (SD = 10.19) in the
treatment group and 29.04 (SD = 9.60) in the control group.

Among participants who completed both the 7-day and 1-month
questionnaires (\(n = 150\)), both groups improved further between the
two follow-up assessments (treatment: M = -1.46, \(p = .038\); control:
M = -2.58, \(p < .001\)), and the additional change did not differ by
condition (\(p = .259\)).
\subsubsection{Mediation}\label{mediation}

\begin{figure}[htbp]

\centering{

\includegraphics[width=0.78\linewidth,height=\textheight,keepaspectratio]{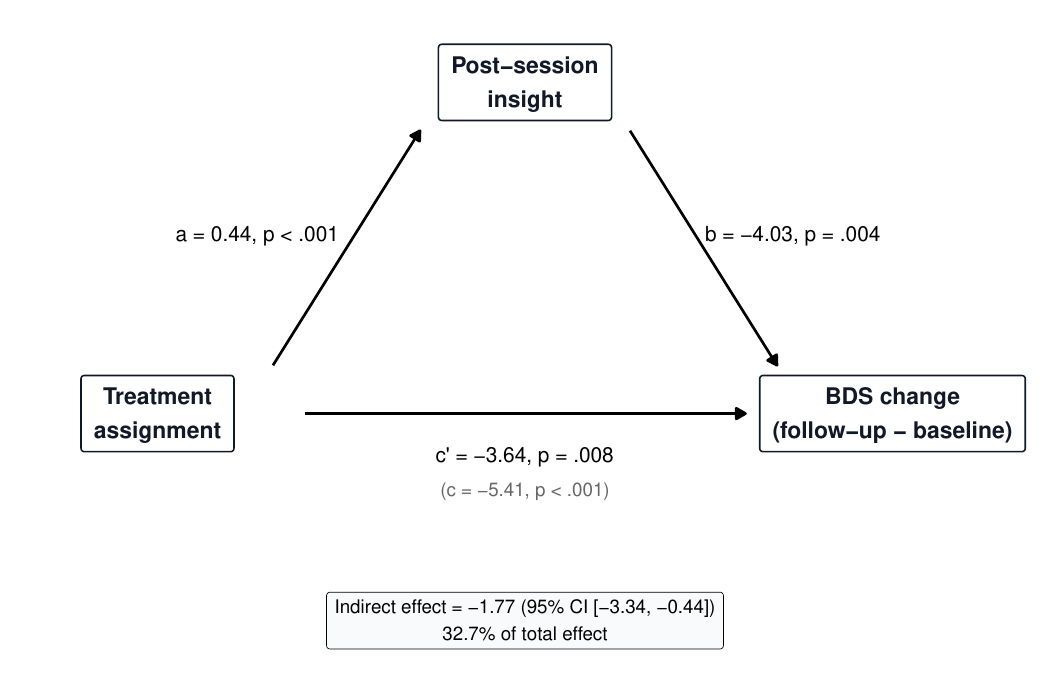}

}

\caption{\label{fig-mediation}Exploratory mediation model for
post-session insight (\emph{N} = 167 participants with both post-session
insight and 7-day BDS follow-up data). Treatment increased the
probability of reporting a post-session insight, and insight was
associated with greater reduction in breakup distress; the remaining
direct effect of treatment was reduced but still significant, consistent
with partial mediation.}

\end{figure}

In further exploratory analyses, post-session insight partially mediated the
treatment effect (Figure~\ref{fig-mediation}). Treatment assignment predicted higher insight reports
(\(B = 0.44\), \(p < .001\)), and insight predicted greater reduction in
breakup distress while controlling for treatment (\(B = -4.03\),
\(p = .004\)). The indirect effect was -1.77, 95\% bootstrap CI
{[}-3.34, -0.44{]}, corresponding to 32.7\% of the total effect. The
direct effect of treatment remained significant after accounting for
insight (\(B = -3.64\), \(p = .008\)), consistent with partial rather
than full mediation.

Follow-up insight at 7 days showed a smaller indirect result, -0.97,
95\% CI {[}-2.14, -0.06{]}, corresponding to 17.5\% of the total effect.

\subsubsection{Moderation}\label{moderation}

Exploratory moderation analyses examined whether baseline
characteristics influenced the magnitude of the treatment effect. None
of the preregistered moderators reached significance: attachment anxiety
(\(p = .165\)), attachment avoidance (\(p = .664\)), time since breakup
(\(p = .807\)), relationship duration (\(p = .197\)), being in a new
relationship (\(p = .352\)), and contact with the former partner
(\(p = .606\)). Baseline breakup-distress severity also did not reach
significance (\(B = -0.17\), \(SE = 0.10\), \(p = .101\)).

Additional post hoc demographic models suggested a larger treatment
effect among male than female participants (\(B = 7.78\), \(p = .003\))
and among younger participants (\(B = -0.23\) per year, \(p = .044\)).

\subsection{User Experience and Qualitative
Feedback}\label{user-experience-and-qualitative-feedback}

\subsubsection{User Experience}\label{user-experience}

Immediately after the intervention, user experience ratings were
generally favorable. UMUX-Lite ease-of-use scores were high in both
groups (treatment: M = 6.51, SD = 0.88; control: M = 6.37, SD = 0.78),
while treatment participants rated the system higher on perceived
capability (5.97 vs.~5.28), empathy and support (6.26 vs.~4.78), trust
and credibility (5.89 vs.~5.03), and anthropomorphism (5.72 vs.~4.56).
The single-item recommendation rating (0--10) was also higher in the
treatment group (M = 7.92, SD = 2.14) than in the control group
(M = 5.80, SD = 2.42), \(p < .001\), \(d = 0.92\). Treatment participants
also more often reported a sudden post-session insight (61.7\%
vs.~19.3\%, \(p < .001\)), and this difference remained evident at 7-day
follow-up among responders (55.7\% vs.~22.8\%, \(p < .001\)).

\subsubsection{Qualitative Feedback}\label{qualitative-feedback}

Open-ended feedback pointed mainly to practical design priorities rather
than to fundamental concerns about safety or usability. The most common
suggestions concerned reducing chatbot response length and the number of
questions stacked into a single turn, modulating pacing and response
latency, and calibrating tone (with some participants finding the
chatbot's belief-challenging directness uncomfortable and others valuing
it). Selected participant illustrations of these themes appear in
Section~\ref{interpretation}.

\section{Discussion}\label{sec-discussion}

\subsection{Main Findings}\label{main-findings}

This randomized controlled trial found that a single session with
\emph{overit} reduced breakup distress more than an assessment-only
control after 7 days. The effect was consistent across the primary
mixed-model analysis and the completer-based change scores. An
additional exploratory 1-month questionnaire suggested that the
treatment advantage persisted, although in attenuated form, and that
both groups continued to improve after the 7-day follow-up rather than
diverging further. Exploratory mediation analyses further suggested that
post-session insight partially accounted for the initial improvement,
whereas preregistered moderator analyses were largely null and only post
hoc demographic models suggested possible heterogeneity by sex and age.
Participants also evaluated the system positively on ease of use,
empathy, and trust, indicating that a structured therapeutic protocol can still
be experienced as conversational and supportive.

\subsection{Interpretation}\label{interpretation}

These findings suggest that a brief structured chatbot conversation can
produce measurable short-term improvement in breakup distress rather
than functioning only as a supportive or engaging user experience. 

The observed effect size (\(d = -0.70\)) exceeds what most comparable
digital interventions have reported. A meta-analysis of 30 mHealth
trials with a cognitive reappraisal component found a pooled effect of
SMD = 0.34 across mental health outcomes
\citep{morelloCognitiveReappraisalMHealth2023}, roughly half the
magnitude observed here, though different outcomes were measured. Among
LLM-based interventions specifically, Socrates 2.0 reported effect sizes
of \(d = 0.25\)--\(0.33\) for symptom reductions via cognitive
reappraisal across a four-week feasibility study
\citep{heldAIFacilitatedCognitiveReappraisal2025}, and Therabot, the
first RCT of a fully generative AI therapy chatbot, reduced depression
symptoms by 39\% over four weeks
\citep{heinzRandomizedTrialGenerative2025}. A multi-session CBT-based
grief app for bereaved parents produced only small effects on grief and
PTSD symptoms \citep{sveenMyGriefApp2025}. An umbrella review of 415
single-session intervention trials found that 83\% reported significant
effects, with a pooled SMD of \(-0.25\)
\citep{schleiderSingleSessionInterventionsMental2025}, roughly a third
of the present effect. These comparisons should be read cautiously:
breakup distress is a narrower and potentially less stable target than
clinical depression or prolonged grief and may be more responsive to
brief intervention, and effect-size estimates from a single small RCT
against an inert control are typically inflated relative to estimates
from active-controlled or replicated trials. Still, a single
conversation producing effects in the range of multi-week programs is
worth taking seriously.

The single-session format also sidesteps a persistent problem in digital
mental health. Attrition rates in eHealth interventions can exceed
50\% \citep{eysenbachLawAttrition2005}, and most users of multi-session
programs complete only a fraction of the intended content
\citep{christensenAdherenceInternetInterventions2009, donkinAdherenceETherapies2011}.
An effective single-session intervention eliminates this problem
entirely. Whether the effect lasts is a separate question: a study of 12
digital single-session interventions for depression found that while
nearly all produced immediate improvements, only 2 of 12 sustained
effects at four weeks
\citep{kaveladzeCrowdsourcedMegastudy12Digital2026}. The exploratory
1 month data in the present study suggest that the treatment advantage
persisted in attenuated form, but both groups continued to improve, and
the gap did not widen. The treatment group still maintained their decreased distress level, but the control participants natural recovery trajectories also led to improvements in breakup distress.  Between day 7 and day 36, the treatment group dropped only 1.46 further points (p = .038) while controls dropped 2.58 (p < .001), with no significant difference between conditions (p = .259). The treatment group had less room to keep improving (mean BDS at 26.6 vs. 32.2), and controls continued to improve at a slower but still relevant rate. One reading is that the chatbot compressed the recovery timeline rather than producing a different endpoint: treatment participants reached in 7 days a level of distress that controls were still approaching at day 36. Whether full convergence between groups eventually occurs cannot be determined from a 36-day window.

The exploratory mediation analyses are consistent with the
intervention's theoretical framing, but they should be interpreted
cautiously. Post-session insight partially mediated the treatment
effect, and follow-up insight showed a smaller indirect effect. This
pattern is compatible with an account, where memory reconsolidation played a role, in line with findings by \citet{speerFindingPositiveMeaning2021} in which
the conversation helps users reactivate a painful memory or memory-based belief, encounter a
contradictory interpretation and emotional state, and experience a shift in meaning
\citep{laneMemoryReconsolidationEmotional2015, eckerMemoryReconsolidationUnderstood2015, sevensterPredictionErrorGoverns2013}.
At the same time, the study was not designed to demonstrate
reconsolidation as a mechanism in an experimental way. The
insight item is only a rough proxy for the kind of mismatch or
prediction error that reconsolidation theories describe, and the
remaining direct effect of treatment suggests that much of the observed
benefit was not explained by this measure alone. Yet, participants'
open-ended descriptions of their insights were often highly emotional updates of maladaptive beliefs, rather than reports of just feeling better or having vented. One
described realizing ``I had been carrying the blame for my failed
relationship instead of recognizing that I did everything the best I
could and I was failed by someone I trusted,'' while another reported
that the chatbot ``related the loneliness to the wall that I had put up
--- that I was self-sabotaging.'' Another participant wrote: ``I was
really surprised. it gave me insights that I just couldn't see myself. I
kind of thought I knew why I was being like I was but I didn't.'' These
sound like actual belief or schema revisions, which may lead to positive changes that remain stable over time.

There remains the question of whether reconsolidation theory offers the
most appropriate explanatory framework for the observed changes, or
whether standard cognitive reappraisal is sufficient. The two accounts
make different predictions about durability. Reconsolidation predicts
lasting change because the updated memory trace replaces the original
\citep{naderReconsolidationDynamicNature2015}, while standard
reappraisal reinterprets the situation without modifying the stored
memory \citep{grossAntecedentResponseFocusedEmotion1998}, so the
original emotional response may re-emerge when the reappraisal is no
longer actively applied. The 7-day follow-up is too short to distinguish between these predictions. At 1 month, treatment-group scores did not rebound but continued to drop slightly; the narrowing gap with controls reflects continued improvement in controls rather than fading of the treatment effect. This is consistent with the durable change that reconsolidation predicts, but since cognitive restructuring interventions can also produce lasting effects \citep{shurickDurableEffectsCognitive2012}, the result pattern is compatible with both accounts.

Longer follow-ups at 3 and 6 months would be needed to
test whether the effect shows the persistence signature expected of
genuine memory updating. Reconsolidation also requires that the target
emotional memory is genuinely reactivated, not merely talked about:
\citet{laneMemoryReconsolidationEmotional2015} argue that new emotional
experiences, not enhanced cognitive understanding alone, drive lasting
change through this mechanism. The qualitative data suggest that the
chatbot conversation did produce emotional engagement --- participants
reported being moved in ways they had not expected, describing the
experience as intense or noting that they needed time to process
afterward --- but the present design cannot confirm whether this
engagement met the threshold for reconsolidation rather than reflecting
the ordinary emotional activation that any structured reflection on a
painful topic would produce \citep{speerFindingPositiveMeaning2021}.

Other mechanisms are also plausible. A structured opportunity to narrate
the breakup, reflect on its meaning, and receive empathic responses may
itself reduce distress through disclosure, meaning making, or
nonspecific support processes
\citep{frattaroliExperimentalDisclosureModerators2006, neimeyerReconstructingMeaningBereavement2011}.
A defensible claim is that a conversational sequence inspired by the memory reconsolidation literature — built around belief reactivation, reinterpretation, and integration — appears useful for short-term breakup recovery. In general, leveraging the findings of memory reconsolidation literature has been useful in the ideation and design phase of the therapeutic agent, because it allowed us to derive concrete design choices — what to elicit first, when to introduce a mismatch, how to consolidate the new interpretation — from a coherent theoretical sequence rather than from intuition alone.

The user-experience results further suggest that this therapeutic sequence
did not feel too rigid for participants. They generally rated the system as
easy to use, empathic, and trustworthy, even though the chatbot followed
a constrained conversation structure enforced by the dual-call
architecture described in Section~\ref{system-architecture}. This matters
because earlier mental health chatbots often faced a tradeoff between
protocol adherence and conversational naturalness
\citep{abd-alrazaqEffectivenessSafetyUsing2020, chaudhryUserPerceptionsExperiences2024,schimpfSupportingEffectiveGoal2026}.
The open-ended feedback in the current analysis snapshot likewise points
to pragmatic design priorities around reducing response length,
modulating pacing, and calibrating chatbot directness. These are
practical rather than theoretical issues, but they are likely important
for whether users remain engaged long enough to benefit from the
intervention.

Exploratory moderation findings were limited. The preregistered
attachment and breakup-context measures did not reliably moderate the
treatment effect, and baseline distress also did not emerge as a
reliable moderator in the updated Python analysis. Post hoc demographic
models suggested a larger treatment effect among male than female
participants (mean BDS change of -14.7 vs.~-3.3 for treatment and
control males, respectively) compared with female participants (-7.4
vs.~-3.8), and a stronger response among younger participants. These
findings raise the possibility that a low-barrier digital intervention
may be particularly well suited to groups who otherwise have fewer
opportunities for structured emotional disclosure
\citep{dindiaSexDifferencesSelfdisclosure1992}. However, the demographic moderation results were based on
unequal and relatively small subgroup cell sizes (21 treatment males
vs.~30 control males among completers) and should be treated as
hypothesis-generating. The main contribution of the study remains the
primary treatment effect, not the moderator profile.

\subsection{Limitations and Future Work}\label{limitations}

Several limitations constrain the strength of the conclusions. First,
the control condition was assessment-only. This makes it impossible to
separate the specific contribution of the reconsolidation-informed
reappraisal sequence from nonspecific factors such as self-disclosure,
time spent reflecting on the breakup, expectancy, or simply having a
responsive conversation with an AI system
\citep{abd-alrazaqEffectivenessSafetyUsing2020}. The treatment condition
was also confounded with personalized score feedback. Treatment
participants saw their breakup distress score as part of the in-app
experience, whereas control participants did not. This asymmetry may
have introduced measurement reactivity if score exposure prompted
self-reflection or anchored subsequent self-reports
\citep{postonMetaanalysisPsychologicalAssessment2010, frenchReactivityMeasurementHealth2010}.
However, the exploratory 1-month follow-up was administered through
external survey software in which no score feedback was displayed, and
treatment participants' average distress scores continued to improve
modestly from the 7-day to the 1-month assessment rather than rebounding.
If reactivity to the displayed score were the primary driver of
improvement, one would expect scores to rebound once that cue was
removed, which was not observed. This does not rule out reactivity as a
contributing factor at the 7-day follow-up, but it suggests that the
treatment effect was not solely an artifact of score exposure. A stronger
next test would compare \emph{overit} against an active
supportive-chatbot control and, ideally, against a chatbot that offers
reflection without the structured belief-challenging / alternative-scenario sequence.

Second, the exploratory mediation analysis relied on a dichotomous
insight item as a proxy for the hypothesized mismatch experience. A
continuous measure of insight quality or magnitude would have provided
more statistical power and a more nuanced estimate of the indirect
effect; the present binary operationalization likely attenuates the
indirect-effect estimate and cannot capture gradations in the
reappraisal process. Future work should incorporate multi-item
continuous measures of insight and cognitive change to test the
mediating pathway more precisely.

Third, the intervention itself was limited to a single session. The
present design therefore cannot tell us whether the observed benefit
reflects the full potential of the approach or whether repeated
check-ins with the chatbot would produce larger, more durable, or more
stable effects over time. The additional exploratory 1-month
questionnaire suggests that some advantage may persist after the initial
session, but that later wave was not preregistered, was restricted to
participants who had already completed the day-0 post-session survey
(biasing it toward more engaged participants), and the gap between
treatment and control noticeably shrank. The mixed-effects models retained incomplete
cases through maximum likelihood estimation, but the estimates still
depend on assumptions about missingness. If dropout was systematically
related to poor response or dissatisfaction, the treatment effect may be
overstated.

Fourth, the study relied entirely on self-report and on a convenience
sample recruited through Prolific. The mean time since breakup of approximately 17.8 months also means the intervention was not primarily tested in the acute post-breakup phase. Participants were likely more
comfortable with digital tools than the broader population, and the
TestFlight-based delivery restricted the sample to iPhone users
\citep{paolacciInsideTurkUnderstanding2014, choDemographicImbalancesBringYourOwnDevice2022}.
The sample was also demographically narrow relative to the range of
people who experience breakup distress. These features limit
generalizability and mean that the findings should be treated as
evidence of initial efficacy rather than broad real-world effectiveness.

These limitations point toward several lines of future work. First, and
most important for the theoretical claims, the reconsolidation account
requires studies that manipulate reactivation and timing directly rather
than inferring mechanism from post-session insight reports. Once a
memory is reactivated, it is thought to enter a labile state lasting
roughly a few hours, during which new input can update the original
trace before it restabilizes
\citep{naderReconsolidationDynamicNature2015, eckerMemoryReconsolidationUnderstood2015}.
The standard reconsolidation paradigm exploits this window by comparing
an intervention delivered within the post-reactivation window against
both a no-reactivation control and a delayed condition in which the
same intervention is delivered outside this window
\citep{schillerPreventingReturnFear2010, sevensterPredictionErrorGoverns2013}.
Only when the within-window group shows durable change while the delayed
group does not can the effect be attributed to reconsolidation rather
than to standard learning or reappraisal. Applying this paradigm to the
digital context, with follow-ups at 1, 3, and 6 months, would test
whether the present effects depend on reactivation timing or reflect
only cognitive change that would work regardless of when it is
delivered.

Second, comparing \emph{overit} against a chatbot that provides empathic
support without the rest of the protocol would test whether the
structured reappraisal components specifically drive the effect or
whether a supportive AI conversation is sufficient on its own. The
present assessment-only control cannot separate the contribution of the
reconsolidation-informed structure from nonspecific therapeutic factors.

Third, neither the dual-call architecture nor the reappraisal sequence
is specific to breakup distress. The same conversational structure ---
reactivating a concrete maladaptive belief, generating a contradictory
interpretation, and checking whether the original emotional response has
shifted --- could in principle be applied to other forms of
belief-driven distress, including grief, workplace stress, or
dysfunctional schemas. If the mechanism turns out to be reconsolidation,
the approach should work wherever a specific emotional memory can be
reactivated and contradicted; if it turns out to be standard cognitive
reappraisal, it should generalize to any domain where cognitive
restructuring is effective. The present study offers early evidence that
this format produces short-term change for one specific stressor.
Whether it generalizes, and whether the mechanism is the one it was
designed to exploit, remain open.

\section{Conclusion}\label{sec-conclusion}

The present study shows that a single structured AI chatbot session can
reduce breakup distress at 7-day follow-up relative to an
assessment-only control, with exploratory evidence that a smaller
treatment advantage remained detectable 1 month later. Exploratory
mediation findings are consistent with, but do not prove, a
reconsolidation-informed mechanism; whether the effect reflects genuine
memory updating or standard cognitive reappraisal cannot be resolved
without reactivation-timing controls and longer follow-ups. Breakup
distress appears to be a viable target for brief, low-barrier
conversational intervention, and the underlying architecture ---
structured belief reactivation, reinterpretation, and integration
delivered through the dual-call system described in
Section~\ref{system-architecture} --- is not specific to breakup
distress and could in principle be applied to other forms of
belief-driven suffering. If future studies can replicate the effect
against active controls, isolate the mechanism more precisely, and test
the approach across different stressors, structured chatbots may offer a
scalable way to provide timely support for common forms of distress that
are often left untreated.

\clearpage
\bibliographystyle{unsrtnat}
\bibliography{bibliography.bib}

@inproceedings{schimpfSupportingEffectiveGoal2026,
  title     = {Supporting Effective Goal Setting with {LLM-Based} Chatbots},
  author    = {Schimpf, Michel and Maier, Sebastian and Wyrowski, Anton and Christoforakos, Lara and Feuerriegel, Stefan and Bohn{\'e}, Thomas},
  booktitle = {Proceedings of the 2026 CHI Conference on Human Factors in Computing Systems (CHI '26)},
  year      = {2026},
  address   = {Barcelona, Spain},
  publisher = {ACM},
  note      = {To appear}
}

@misc{schimpfAIAssistedGoalSettingImproves2026,
  title         = {{AI-Assisted Goal Setting Improves Goal Progress Through Social Accountability}},
  author        = {Schimpf, Michel and Voigt, Julian and Bohn{\'e}, Thomas},
  year          = {2026},
  eprint        = {2603.17887},
  archivePrefix = {arXiv},
  primaryClass  = {cs.HC},
  url           = {https://arxiv.org/abs/2603.17887}
}

@article{dindiaSexDifferencesSelfdisclosure1992,
  title = {Sex Differences in Self-Disclosure: {{A}} Meta-Analysis},
  author = {Dindia, Kathryn and Allen, Mike},
  year = {1992},
  journal = {Psychological Bulletin},
  volume = {112},
  number = {1},
  pages = {106--124},
  doi = {10.1037/0033-2909.112.1.106}
}

@article{grossAntecedentResponseFocusedEmotion1998,
  title = {Antecedent- and Response-Focused Emotion Regulation: {{Divergent}} Consequences for Experience, Expression, and Physiology},
  author = {Gross, James J.},
  year = {1998},
  journal = {Journal of Personality and Social Psychology},
  volume = {74},
  number = {1},
  pages = {224--237},
  doi = {10.1037/0022-3514.74.1.224}
}

@article{boelenCognitiveBehavioralConceptualization2006,
  title = {A Cognitive-Behavioral Conceptualization of Complicated Grief.},
  volume = {13},
  ISSN = {0969-5893},
  url = {http://dx.doi.org/10.1111/j.1468-2850.2006.00013.x},
  DOI = {10.1111/j.1468-2850.2006.00013.x},
  number = {2},
  journal = {Clinical Psychology: Science and Practice},
  publisher = {American Psychological Association (APA)},
  author = {Boelen,  Paul A. and van den Hout,  Marcel A. and van den Bout,  Jan},
  year = {2006},
  pages = {109–128}
}

@article{lewisImplementingMeasurementBased2019,
  title = {Implementing {{Measurement-Based Care}} in {{Behavioral Health}}: {{A Review}}},
  author = {Lewis, Cara C. and Boyd, Maren and Puspitasari, Ajeng and Navarro, Elena and Howard, Jadyn and Kassab, Hannah and Hoffman, Mylynn and Scott, Kelli and Lyon, Aaron and Douglas, Suzanne and Simon, Greg and Kroenke, Kurt},
  year = {2019},
  journal = {JAMA Psychiatry},
  volume = {76},
  number = {3},
  pages = {324--335},
  doi = {10.1001/jamapsychiatry.2018.3329}
}

@article{heinzRandomizedTrialGenerative2025,
  title = {Randomized Trial of a Generative {{AI}} Chatbot for Mental Health Treatment},
  author = {Heinz, Michael V. and Mackin, Daniel M. and Trudeau, Brianna M. and Bhattacharya, Sukanya and Wang, Yinzhou and Banta, Haley A. and Jewett, Abi D. and Salzhauer, Abigail J. and Griffin, Tess Z. and Jacobson, Nicholas C.},
  year = {2025},
  journal = {NEJM AI},
  volume = {2},
  number = {4},
  doi = {10.1056/AIoa2400802}
}

@article{schleiderSingleSessionInterventionsMental2025,
  title = {Single-Session Interventions for Mental Health Problems and Service Engagement: {{Umbrella}} Review of Systematic Reviews and Meta-Analyses},
  author = {Schleider, Jessica L. and Zapata, Juan Pablo and Rapoport, Andy and Wescott, Annie and Ghosh, Arka and Kaveladze, Benji and Szkody, Erica and Ahuvia, Isaac L.},
  year = {2025},
  journal = {Annual Review of Clinical Psychology},
  volume = {21},
  number = {1},
  pages = {279--303},
  doi = {10.1146/annurev-clinpsy-081423-025033}
}

@article{kaveladzeCrowdsourcedMegastudy12Digital2026,
  title = {A Crowdsourced Megastudy of 12 Digital Single-Session Interventions for Depression in {{US}} Adults},
  author = {Kaveladze, Benjamin and Voelkel, Jan and Stagnaro, Michael and others},
  year = {2026},
  journal = {Nature Human Behaviour},
  volume = {10},
  doi = {10.1038/s41562-026-02415-6}
}

@article{paolacciInsideTurkUnderstanding2014,
  title = {Inside the {{Turk}}: {{Understanding Mechanical Turk}} as a Participant Pool},
  author = {Paolacci, Gabriele and Chandler, Jesse},
  year = {2014},
  journal = {Current Directions in Psychological Science},
  volume = {23},
  number = {3},
  pages = {184--188},
  doi = {10.1177/0963721414531598}
}

@article{frattaroliExperimentalDisclosureModerators2006,
  title = {Experimental Disclosure and Its Moderators: {{A}} Meta-Analysis},
  author = {Frattaroli, Joanne},
  year = {2006},
  journal = {Psychological Bulletin},
  volume = {132},
  number = {6},
  pages = {823--865},
  doi = {10.1037/0033-2909.132.6.823}
}

@inproceedings{jiangFollowBenchMultilevelFinegrained2024,
  title = {{{FollowBench}}: A Multi-Level Fine-Grained Constraints Following Benchmark for Large Language Models},
  author = {Jiang, Yuxin and Wang, Yufei and Zeng, Xingshan and Zhong, Wanjun and Li, Liangyou and Mi, Fei and Shang, Lifeng and Jiang, Xin and Liu, Qun and Wang, Wei},
  booktitle = {Proceedings of the 62nd Annual Meeting of the Association for Computational Linguistics (Volume 1: Long Papers)},
  year = {2024},
  pages = {4667--4688},
  publisher = {Association for Computational Linguistics},
  address = {Bangkok, Thailand},
  doi = {10.18653/v1/2024.acl-long.257}
}

@article{abd-alrazaqEffectivenessSafetyUsing2020,
  title = {Effectiveness and {{Safety}} of {{Using Chatbots}} to {{Improve Mental Health}}: {{Systematic Review}} and {{Meta-Analysis}}},
  shorttitle = {Effectiveness and {{Safety}} of {{Using Chatbots}} to {{Improve Mental Health}}},
  author = {{Abd-Alrazaq}, Alaa Ali and Rababeh, Asma and Alajlani, Mohannad and Bewick, Bridgette M and Househ, Mowafa},
  year = {2020},
  month = jul,
  journal = {Journal of Medical Internet Research},
  volume = {22},
  number = {7},
  pages = {e16021},
  issn = {1439-4456},
  doi = {10.2196/16021},
  urldate = {2026-02-12},
  pmcid = {PMC7385637},
  pmid = {32673216}
}

@article{brennerMeasuringThoughtContent2015,
  title = {Measuring Thought Content Valence after a Breakup: {{Development}} of the {{Positive}} and {{Negative Ex-Relationship Thoughts}} ({{PANERT}}) Scale.},
  shorttitle = {Measuring Thought Content Valence after a Breakup},
  author = {Brenner, Rachel E. and Vogel, David L.},
  year = {2015},
  month = jul,
  journal = {Journal of Counseling Psychology},
  volume = {62},
  number = {3},
  pages = {476--487},
  issn = {1939-2168, 0022-0167},
  doi = {10.1037/cou0000073},
  urldate = {2025-10-04},
  langid = {english}
}

@article{chaudhryUserPerceptionsExperiences2024,
  title = {User Perceptions and Experiences of an {{AI-driven}} Conversational Agent for Mental Health Support},
  author = {Chaudhry, Beenish Moalla and Debi, Happy Rani},
  year = {2024},
  journal = {mHealth},
  volume = {10},
  pages = {21},
  doi = {10.21037/mhealth-23-55},
  pmcid = {PMC11304096},
  pmid = {39114473}
}

@article{eckerMemoryReconsolidationUnderstood2015,
  title = {Memory Reconsolidation Understood and Misunderstood},
  author = {Ecker, Bruce},
  year = {2015},
  month = jan,
  journal = {International Journal of Neuropsychotherapy},
  volume = {3},
  number = {1},
  pages = {2--46},
  issn = {22027653},
  doi = {10.12744/ijnpt.2015.0002-0046},
  urldate = {2025-10-02},
  langid = {english}
}

@article{fieldBreakupDistressUniversity2009,
  title = {Breakup Distress in University Students},
  author = {Field, Tiffany and Diego, Miguel and Pelaez, Martha and Deeds, Osvelia and Delgado, Jeannette},
  year = {2009},
  journal = {Adolescence},
  volume = {44},
  number = {176},
  pages = {705--727},
  issn = {0001-8449},
  langid = {english},
  pmid = {20432597}
}

@article{gehlAttachmentBreakupDistress2024,
  title = {Attachment and {{Breakup Distress}}: {{The Mediating Role}} of {{Coping Strategies}}},
  shorttitle = {Attachment and {{Breakup Distress}}},
  author = {Gehl, Kristin and Brassard, Audrey and Dugal, Caroline and Lefebvre, Audrey-Ann and Daigneault, Isabelle and Francoeur, Audrey and Lecomte, Tania},
  year = {2024},
  month = feb,
  journal = {Emerging Adulthood (Print)},
  volume = {12},
  number = {1},
  pages = {41--54},
  issn = {2167-6968},
  doi = {10.1177/21676968231209232},
  urldate = {2026-02-07},
  pmcid = {PMC10727987},
  pmid = {38124712}
}

@article{heldAIFacilitatedCognitiveReappraisal2025,
  title = {{{AI-Facilitated Cognitive Reappraisal}} via {{Socrates}} 2.0: {{Mixed Methods Feasibility Study}}},
  shorttitle = {{{AI-Facilitated Cognitive Reappraisal}} via {{Socrates}} 2.0},
  author = {Held, Philip and Pridgen, Sarah A. and Szoke, Daniel R. and Chen, Yaozhong and Akhtar, Zuhaib and Amin, Darpan},
  year = {2025},
  journal = {JMIR Mental Health},
  volume = {12},
  pages = {e80461},
  publisher = {JMIR Publications Toronto, Canada},
  urldate = {2026-02-07}
}

@article{laneMemoryReconsolidationEmotional2015,
  title = {Memory Reconsolidation, Emotional Arousal, and the Process of Change in Psychotherapy: {{New}} Insights from Brain Science},
  shorttitle = {Memory Reconsolidation, Emotional Arousal, and the Process of Change in Psychotherapy},
  author = {Lane, Richard D. and Ryan, Lee and Nadel, Lynn and Greenberg, Leslie},
  year = {2015},
  month = jan,
  journal = {Behavioral and Brain Sciences},
  volume = {38},
  pages = {e1},
  issn = {0140-525X, 1469-1825},
  doi = {10.1017/S0140525X14000041},
  urldate = {2026-02-07},
  langid = {english}
}

@article{langeslagDownregulationLoveFeelings2018,
  title = {Down-Regulation of Love Feelings after a Romantic Break-up: {{Self-report}} and Electrophysiological Data},
  shorttitle = {Down-Regulation of Love Feelings after a Romantic Break-Up},
  author = {Langeslag, Sandra J. E. and Sanchez, Michelle E.},
  year = {2018},
  month = may,
  journal = {Journal of Experimental Psychology. General},
  volume = {147},
  number = {5},
  pages = {720--733},
  issn = {1939-2222},
  doi = {10.1037/xge0000360},
  langid = {english},
  pmid = {28857575}
}

@inproceedings{lewisUMUXLITEWhenTheres2013,
  title = {{{UMUX-LITE}}: When There's No Time for the {{SUS}}},
  shorttitle = {{{UMUX-LITE}}},
  booktitle = {Proceedings of the {{SIGCHI Conference}} on {{Human Factors}} in {{Computing Systems}}},
  author = {Lewis, James R. and Utesch, Brian S. and Maher, Deborah E.},
  year = {2013},
  month = apr,
  series = {{{CHI}} '13},
  pages = {2099--2102},
  publisher = {Association for Computing Machinery},
  address = {New York, NY, USA},
  doi = {10.1145/2470654.2481287},
  urldate = {2026-02-06},
  isbn = {978-1-4503-1899-0}
}

@article{morelloCognitiveReappraisalMHealth2023,
  title = {Cognitive Reappraisal in {{mHealth}} Interventions to Foster Mental Health in Adults: A Systematic Review and Meta-Analysis},
  shorttitle = {Cognitive Reappraisal in {{mHealth}} Interventions to Foster Mental Health in Adults},
  author = {Morello, Karolina and Sch{\"a}fer, Sarah K. and Kunzler, Angela M. and Priesterroth, Lilli-Sophie and T{\"u}scher, Oliver and Kubiak, Thomas},
  year = {2023},
  month = oct,
  journal = {Frontiers in Digital Health},
  volume = {5},
  publisher = {Frontiers},
  issn = {2673-253X},
  doi = {10.3389/fdgth.2023.1253390},
  urldate = {2026-02-07},
  langid = {english}
}

@article{naderReconsolidationDynamicNature2015,
  title = {Reconsolidation and the {{Dynamic Nature}} of {{Memory}}},
  author = {Nader, Karim},
  year = {2015},
  month = oct,
  journal = {Cold Spring Harbor Perspectives in Biology},
  volume = {7},
  number = {10},
  pages = {a021782},
  issn = {1943-0264},
  doi = {10.1101/cshperspect.a021782},
  urldate = {2026-02-07},
  pmcid = {PMC4588064},
  pmid = {26354895}
}

@incollection{neimeyerReconstructingMeaningBereavement2011,
  title = {Reconstructing {{Meaning}} in {{Bereavement}}},
  booktitle = {Handbook of {{Psychotherapy}} in {{Cancer Care}}},
  author = {Neimeyer, Robert A.},
  year = {2011},
  pages = {247--257},
  publisher = {John Wiley \& Sons, Ltd},
  doi = {10.1002/9780470975176.ch21},
  urldate = {2026-02-07},
  chapter = {21},
  copyright = {Copyright {\copyright} 2011 John Wiley \& Sons, Ltd},
  isbn = {978-0-470-97517-6},
  langid = {english}
}

@article{sorinLargeLanguageModels2024,
  title = {Large {{Language Models}} and {{Empathy}}: {{Systematic Review}}},
  shorttitle = {Large {{Language Models}} and {{Empathy}}},
  author = {Sorin, Vera and Brin, Dana and Barash, Yiftach and Konen, Eli and Charney, Alexander and Nadkarni, Girish and Klang, Eyal},
  year = {2024},
  month = dec,
  journal = {Journal of Medical Internet Research},
  volume = {26},
  number = {1},
  pages = {e52597},
  publisher = {JMIR Publications Inc., Toronto, Canada},
  doi = {10.2196/52597},
  urldate = {2026-02-12},
  langid = {english}
}

@article{speerFindingPositiveMeaning2021,
  title = {Finding Positive Meaning in Memories of Negative Events Adaptively Updates Memory},
  author = {Speer, Megan E. and Ibrahim, Sandra and Schiller, Daniela and Delgado, Mauricio R.},
  year = {2021},
  month = nov,
  journal = {Nature Communications},
  volume = {12},
  number = {1},
  pages = {6601},
  publisher = {Nature Publishing Group},
  issn = {2041-1723},
  doi = {10.1038/s41467-021-26906-4},
  urldate = {2025-08-10},
  copyright = {2021 The Author(s)},
  langid = {english}
}

@article{stoverMetaanalysisCognitiveReappraisal2024,
  title = {A Meta-Analysis of Cognitive Reappraisal and Personal Resilience},
  author = {Stover, Alexander D. and Shulkin, Josh and Lac, Andrew and Rapp, Timothy},
  year = {2024},
  month = jun,
  journal = {Clinical Psychology Review},
  volume = {110},
  pages = {102428},
  issn = {0272-7358},
  doi = {10.1016/j.cpr.2024.102428},
  urldate = {2025-10-05}
}

@article{sveenMyGriefApp2025,
  title = {My Grief App for Prolonged Grief in Bereaved Parents: A Randomised Waitlist-Controlled Trial},
  shorttitle = {My Grief App for Prolonged Grief in Bereaved Parents},
  author = {Sveen, Josefin and Eisma, Maarten C. and Boelen, Paul A. and Arnberg, Filip K. and Eklund, Rakel},
  year = {2025},
  month = jul,
  journal = {Cognitive Behaviour Therapy},
  volume = {54},
  number = {4},
  pages = {514--530},
  publisher = {Routledge},
  issn = {1650-6073},
  doi = {10.1080/16506073.2024.2429068},
  urldate = {2026-02-07},
  pmid = {39540459}
}

@article{shurickDurableEffectsCognitive2012,
  title = {Durable Effects of Cognitive Restructuring on Conditioned Fear},
  author = {Shurick, Ashley A. and Hamilton, Jeffrey R. and Harris, Lasana T. and Roy, Amy K. and Gross, James J. and Phelps, Elizabeth A.},
  year = {2012},
  journal = {Emotion},
  volume = {12},
  number = {6},
  pages = {1393--1397},
  doi = {10.1037/a0029143}
}

@article{verhallenRomanticRelationshipBreakup2019,
  title = {Romantic Relationship Breakup: {{An}} Experimental Model to Study Effects of Stress on Depression (-like) Symptoms},
  shorttitle = {Romantic Relationship Breakup},
  author = {Verhallen, Anne M. and Renken, Remco J. and Marsman, Jan-Bernard C. and Ter Horst, Gert J.},
  editor = {Blanch, Angel},
  year = {2019},
  month = may,
  journal = {PLOS ONE},
  volume = {14},
  number = {5},
  pages = {e0217320},
  issn = {1932-6203},
  doi = {10.1371/journal.pone.0217320},
  urldate = {2025-09-29},
  langid = {english}
}

@article{weiExperiencesCloseRelationship2007,
  title = {The {{Experiences}} in {{Close Relationship Scale}} ({{ECR}})-Short Form: Reliability, Validity, and Factor Structure},
  shorttitle = {The {{Experiences}} in {{Close Relationship Scale}} ({{ECR}})-Short Form},
  author = {Wei, Meifen and Russell, Daniel W. and Mallinckrodt, Brent and Vogel, David L.},
  year = {2007},
  month = apr,
  journal = {Journal of Personality Assessment},
  volume = {88},
  number = {2},
  pages = {187--204},
  issn = {0022-3891},
  doi = {10.1080/00223890701268041},
  langid = {english},
  pmid = {17437384}
}

@article{mcbainUseGenerativeAI2025,
  title = {Use of Generative {{AI}} for Mental Health Advice among {{US}} Adolescents and Young Adults},
  author = {McBain, Ryan K. and Bozick, Robert and Diliberti, Melissa and Zhang, Li Ang and Zhang, Fang and Burnett, Alyssa and Kofner, Aaron and Rader, Benjamin and Breslau, Joshua and Stein, Bradley D. and Mehrotra, Ateev and Uscher-Pines, Lori and Cantor, Jonathan and Yu, Hao},
  year = {2025},
  journal = {JAMA Network Open},
  volume = {8},
  number = {11},
  pages = {e2542281},
  doi = {10.1001/jamanetworkopen.2025.42281}
}

@article{ayersComparingPhysicianArtificial2023,
  title = {Comparing Physician and Artificial Intelligence Chatbot Responses to Patient Questions Posted to a Public Social Media Forum},
  author = {Ayers, John W. and Poliak, Adam and Dredze, Mark and Leas, Eric C. and Zhu, Zechariah and Kelley, Jessica B. and Faix, Dennis J. and Goodman, Aaron M. and Longhurst, Christopher A. and Hogarth, Michael and Smith, Davey M.},
  year = {2023},
  journal = {JAMA Internal Medicine},
  volume = {183},
  number = {6},
  pages = {589--596},
  doi = {10.1001/jamainternmed.2023.1838}
}

@inproceedings{ouyangTrainingLanguageModels2022,
  title = {Training Language Models to Follow Instructions with Human Feedback},
  author = {Ouyang, Long and Wu, Jeffrey and Jiang, Xu and Almeida, Diogo and Wainwright, Carroll and Mishkin, Pamela and Zhang, Chong and Agarwal, Sandhini and Slama, Katarina and Ray, Alex and Schulman, John and Hilton, Jacob and Kelton, Fraser and Miller, Luke and Simens, Maddie and Askell, Amanda and Welinder, Peter and Christiano, Paul F. and Leike, Jan and Lowe, Ryan},
  year = {2022},
  booktitle = {Advances in Neural Information Processing Systems},
  volume = {35},
  pages = {27730--27744}
}

@inproceedings{sharmaTowardsUnderstandingSycophancy2024,
  title = {Towards Understanding Sycophancy in Language Models},
  author = {Sharma, Mrinank and Tong, Meg and Korbak, Tomasz and Duvenaud, David and Askell, Amanda and Bowman, Samuel R. and Durmus, Esin and Hatfield-Dodds, Zac and Johnston, Scott R. and Kravec, Shauna and Maxwell, Timothy and McCandlish, Sam and Ndousse, Kamal and Rausch, Oliver and Schiefer, Nicholas and Yan, Da and Zhang, Miranda and Perez, Ethan},
  year = {2024},
  booktitle = {Proceedings of the International Conference on Learning Representations},
  doi = {10.48550/arXiv.2310.13548}
}

@article{malmqvistSycophancyLargeLanguage2024,
  title = {Sycophancy in Large Language Models: Causes and Mitigations},
  author = {Malmqvist, Lars},
  year = {2024},
  journal = {arXiv},
  eprint = {2411.15287},
  archiveprefix = {arXiv},
  primaryclass = {cs.CL}
}

@inproceedings{perezDiscoveringLanguageModel2023,
  title = {Discovering Language Model Behaviors with Model-Written Evaluations},
  author = {Perez, Ethan and Ringer, Sam and Lukosiute, Kamile and Nguyen, Karina and Chen, Edwin and Heiner, Scott and Pettit, Craig and Olsson, Catherine and Kundu, Sandipan and Kadavath, Saurav and Jones, Andy and Chen, Anna and Mann, Benjamin and Israel, Brian and Seethor, Bryan and McKinnon, Cameron and Olah, Christopher and Yan, Da and Amodei, Daniela and Amodei, Dario and others},
  year = {2023},
  booktitle = {Findings of the Association for Computational Linguistics: {{ACL}} 2023},
  pages = {13387--13434},
  doi = {10.18653/v1/2023.findings-acl.847}
}

@inproceedings{mooreExpressingStigmaInappropriate2025,
  title = {Expressing Stigma and Inappropriate Responses Prevents {{LLMs}} from Safely Replacing Mental Health Providers},
  author = {Moore, Jared and Grabb, Declan and Agnew, William and Klyman, Kevin and Chancellor, Stevie and Ong, Desmond C. and Haber, Nick},
  year = {2025},
  booktitle = {Proceedings of the 2025 {{ACM Conference}} on {{Fairness}}, {{Accountability}}, and {{Transparency}}},
  pages = {599--627},
  doi = {10.1145/3715275.3732039}
}

@book{beckCognitiveTherapyDepression1979,
  title = {Cognitive Therapy of Depression},
  author = {Beck, Aaron T. and Rush, A. John and Shaw, Brian F. and Emery, Gary},
  year = {1979},
  publisher = {Guilford Press},
  address = {New York}
}

@article{hofmannEfficacyCognitiveBehavioral2012,
  title = {The Efficacy of Cognitive Behavioral Therapy: {{A}} Review of Meta-Analyses},
  author = {Hofmann, Stefan G. and Asnaani, Anu and Vonk, Imke J. J. and Sawyer, Alice T. and Fang, Angela},
  year = {2012},
  journal = {Cognitive Therapy and Research},
  volume = {36},
  number = {5},
  pages = {427--440},
  doi = {10.1007/s10608-012-9476-1}
}

@article{liuLostMiddleHow2024,
  title = {Lost in the Middle: {{How}} Language Models Use Long Contexts},
  author = {Liu, Nelson F. and Lin, Kevin and Hewitt, John and Paranjape, Ashwin and Bevilacqua, Michele and Petroni, Fabio and Liang, Percy},
  year = {2024},
  journal = {Transactions of the Association for Computational Linguistics},
  volume = {12},
  pages = {157--173},
  doi = {10.1162/tacl_a_00638}
}

@inproceedings{zhengJudgingLLMasJudge2023,
  title = {Judging {LLM}-as-a-Judge with {MT-Bench} and {Chatbot Arena}},
  author = {Zheng, Lianmin and Chiang, Wei-Lin and Sheng, Ying and Zhuang, Siyuan and Wu, Zhanghao and Zhuang, Yonghao and Lin, Zi and Li, Zhuohan and Li, Dacheng and Xing, Eric and Zhang, Hao and Gonzalez, Joseph E. and Stoica, Ion},
  booktitle = {Advances in Neural Information Processing Systems},
  volume = {36},
  pages = {46595--46623},
  year = {2023},
  publisher = {Curran Associates, Inc.}
}

@article{wasenmullerScriptBasedDialog2024,
  title = {Script-Based Dialog Policy Planning for {LLM}-Powered Conversational Agents: {{A}} Basic Architecture for an ``{AI} Therapist''},
  author = {Wasenm{\"u}ller, Robert and Hilbert, Kevin and Benzm{\"u}ller, Christoph},
  year = {2024},
  journal = {arXiv preprint arXiv:2412.15242},
  doi = {10.48550/arXiv.2412.15242}
}

@misc{auYeungPsychogenicMachineSimulating2025,
  title = {The {{Psychogenic Machine}}: {{Simulating AI Psychosis}}, {{Delusion Reinforcement}} and {{Harm Enablement}} in {{Large Language Models}}},
  author = {{Au Yeung}, Joshua and Dalmasso, Jacopo and Foschini, Luca and Dobson, Richard J. B. and Kraljevic, Zeljko},
  year = {2025},
  month = sep,
  eprint = {2509.10970},
  archiveprefix = {arXiv},
  primaryclass = {cs.CL}
}

@article{boelenNegativeCognitionsEmotional2009,
  title = {Negative Cognitions in Emotional Problems Following Romantic Relationship Break-Ups},
  author = {Boelen, Paul A. and Reijntjes, Albert},
  year = {2009},
  journal = {Stress and Health},
  volume = {25},
  number = {1},
  pages = {11--19},
  doi = {10.1002/smi.1219}
}

@article{slotterWhoAmYou2010,
  title = {Who Am {{I}} without You? {{The}} Influence of Romantic Breakup on the Self-Concept},
  author = {Slotter, Erica B. and Gardner, Wendi L. and Finkel, Eli J.},
  year = {2010},
  journal = {Personality and Social Psychology Bulletin},
  volume = {36},
  number = {2},
  pages = {147--160},
  doi = {10.1177/0146167209352250}
}

@article{ezawaCognitiveRestructuringPsychotherapy2023,
  title = {Cognitive Restructuring and Psychotherapy Outcome: {{A}} Meta-Analytic Review},
  author = {Ezawa, Irene D. and Hollon, Steven D.},
  year = {2023},
  journal = {Psychotherapy},
  volume = {60},
  number = {3},
  pages = {396--406},
  doi = {10.1037/pst0000474}
}

@article{schillerPreventingReturnFear2010,
  title = {Preventing the Return of Fear in Humans Using Reconsolidation Update Mechanisms},
  author = {Schiller, Daniela and Monfils, Marie-H. and Raio, Candace M. and Johnson, David C. and LeDoux, Joseph E. and Phelps, Elizabeth A.},
  year = {2010},
  journal = {Nature},
  volume = {463},
  number = {7277},
  pages = {49--53},
  doi = {10.1038/nature08637}
}

@article{sevensterPredictionErrorGoverns2013,
  title = {Prediction Error Governs Pharmacologically Induced Amnesia for Learned Fear},
  author = {Sevenster, Dieuwke and Beckers, Tom and Kindt, Merel},
  year = {2013},
  journal = {Science},
  volume = {339},
  number = {6121},
  pages = {830--833},
  doi = {10.1126/science.1231357}
}

@book{eckerUnlockingEmotionalBrain2012,
  title = {Unlocking the Emotional Brain: {{Eliminating}} Symptoms at Their Roots Using Memory Reconsolidation},
  author = {Ecker, Bruce and Ticic, Robin and Hulley, Laurel},
  year = {2012},
  publisher = {Routledge},
  address = {New York},
  isbn = {978-0-415-89717-4}
}

@article{eysenbachLawAttrition2005,
  title = {The Law of Attrition},
  author = {Eysenbach, Gunther},
  year = {2005},
  journal = {Journal of Medical Internet Research},
  volume = {7},
  number = {1},
  pages = {e11},
  doi = {10.2196/jmir.7.1.e11}
}

@article{christensenAdherenceInternetInterventions2009,
  title = {Adherence in Internet Interventions for Anxiety and Depression: {{Systematic}} Review},
  author = {Christensen, Helen and Griffiths, Kathleen M. and Farrer, Louise},
  year = {2009},
  journal = {Journal of Medical Internet Research},
  volume = {11},
  number = {2},
  pages = {e13},
  doi = {10.2196/jmir.1194}
}

@article{donkinAdherenceETherapies2011,
  title = {A Systematic Review of the Impact of Adherence on the Effectiveness of e-Therapies},
  author = {Donkin, Liesje and Christensen, Helen and Naismith, Sharon L. and Neal, Bruce and Hickie, Ian B. and Glozier, Nick},
  year = {2011},
  journal = {Journal of Medical Internet Research},
  volume = {13},
  number = {3},
  pages = {e52},
  doi = {10.2196/jmir.1772}
}

@article{frenchReactivityMeasurementHealth2010,
  title = {Reactivity of Measurement in Health Psychology: {{How}} Much of a Problem Is It? {{What}} Can Be Done about It?},
  author = {French, David P. and Sutton, Stephen},
  year = {2010},
  journal = {British Journal of Health Psychology},
  volume = {15},
  number = {3},
  pages = {453--468},
  doi = {10.1348/135910710X492341}
}

@article{perillouxBreakingRomanticRelationships2008,
  title = {Breaking up Romantic Relationships: {{Costs}} Experienced and Coping Strategies Deployed},
  author = {Perilloux, Carin and Buss, David M.},
  year = {2008},
  journal = {Evolutionary Psychology},
  volume = {6},
  number = {1},
  pages = {164--181},
  doi = {10.1177/147470490800600119}
}

@article{postonMetaanalysisPsychologicalAssessment2010,
  title = {Meta-Analysis of Psychological Assessment as a Therapeutic Intervention},
  author = {Poston, John M. and Hanson, William E.},
  year = {2010},
  journal = {Psychological Assessment},
  volume = {22},
  number = {2},
  pages = {203--212},
  doi = {10.1037/a0018679}
}

@article{choDemographicImbalancesBringYourOwnDevice2022,
  title = {Demographic Imbalances Resulting From the Bring-Your-Own-Device Study Design},
  author = {Cho, Peter Jaeho and Yi, Jaehan and Ho, Ethan and Shandhi, Md Mobashir Hasan and Dinh, Yen and Patil, Aneesh and Martin, Leatrice and Singh, Geetika and Bent, Brinnae and Ginsburg, Geoffrey and Smuck, Matthew and Woods, Christopher and Shaw, Ryan and Dunn, Jessilyn},
  year = {2022},
  journal = {JMIR mHealth and uHealth},
  volume = {10},
  number = {4},
  pages = {e29510},
  doi = {10.2196/29510}
}

@article{fieldBreakupDistressLoss2010,
  title = {Breakup Distress and Loss of Intimacy in University Students},
  author = {Field, Tiffany and Diego, Miguel and Pelaez, Martha and Deeds, Osvelia and Delgado, Jeannette},
  year = {2010},
  journal = {Psychology},
  volume = {1},
  number = {3},
  pages = {173--177},
  doi = {10.4236/psych.2010.13023}
}

\clearpage
\appendix
\section*{Appendix}
\addcontentsline{toc}{section}{Appendix}

\section{Methods Details}\label{sec-methods-appendix}

\subsection{Scale Properties}\label{scale-properties}

Table~\ref{tbl-scale-properties} summarizes the psychometric properties
of the multi-item scales used in this study, as reported in their
original validation studies.

\begin{longtable}[]{@{}
  >{\raggedright\arraybackslash}p{(\linewidth - 12\tabcolsep) * \real{0.1429}}
  >{\raggedright\arraybackslash}p{(\linewidth - 12\tabcolsep) * \real{0.1429}}
  >{\raggedright\arraybackslash}p{(\linewidth - 12\tabcolsep) * \real{0.1429}}
  >{\raggedright\arraybackslash}p{(\linewidth - 12\tabcolsep) * \real{0.1429}}
  >{\raggedright\arraybackslash}p{(\linewidth - 12\tabcolsep) * \real{0.1429}}
  >{\raggedright\arraybackslash}p{(\linewidth - 12\tabcolsep) * \real{0.1429}}
  >{\raggedright\arraybackslash}p{(\linewidth - 12\tabcolsep) * \real{0.1429}}@{}}
\caption{Psychometric properties of multi-item scales as reported in
their original validation studies. Internal consistency ranges reflect
values across multiple validation
samples.}\label{tbl-scale-properties}\tabularnewline
\toprule\noalign{}
\begin{minipage}[b]{\linewidth}\raggedright
Scale
\end{minipage} & \begin{minipage}[b]{\linewidth}\raggedright
Reference
\end{minipage} & \begin{minipage}[b]{\linewidth}\raggedright
Items
\end{minipage} & \begin{minipage}[b]{\linewidth}\raggedright
Response Scale
\end{minipage} & \begin{minipage}[b]{\linewidth}\raggedright
Scoring
\end{minipage} & \begin{minipage}[b]{\linewidth}\raggedright
Internal Consistency (\(\alpha\))
\end{minipage} & \begin{minipage}[b]{\linewidth}\raggedright
Test--Retest
\end{minipage} \\
\midrule\noalign{}
\endfirsthead
\toprule\noalign{}
\begin{minipage}[b]{\linewidth}\raggedright
Scale
\end{minipage} & \begin{minipage}[b]{\linewidth}\raggedright
Reference
\end{minipage} & \begin{minipage}[b]{\linewidth}\raggedright
Items
\end{minipage} & \begin{minipage}[b]{\linewidth}\raggedright
Response Scale
\end{minipage} & \begin{minipage}[b]{\linewidth}\raggedright
Scoring
\end{minipage} & \begin{minipage}[b]{\linewidth}\raggedright
Internal Consistency (\(\alpha\))
\end{minipage} & \begin{minipage}[b]{\linewidth}\raggedright
Test--Retest
\end{minipage} \\
\midrule\noalign{}
\endhead
\bottomrule\noalign{}
\endlastfoot
BDS & \citep{fieldBreakupDistressUniversity2009} & 16 & 1--4 & Sum
(16--64) & .91 \citep{fieldBreakupDistressLoss2010} & --- \\
ECR-S Anxiety & \citep{weiExperiencesCloseRelationship2007} & 6 & 1--7 &
Mean (1--7) & .77--.86 & .80 \\
ECR-S Avoidance & \citep{weiExperiencesCloseRelationship2007} & 6 & 1--7
& Mean (1--7) & .78--.88 & .83 \\
UMUX-Lite & \citep{lewisUMUXLITEWhenTheres2013} & 2 & 1--7 & Item-level
& .82--.83 & --- \\
\end{longtable}

\section{Results Details}\label{results-details}

\subsection{Baseline Characteristics}\label{baseline-characteristics}

Table~\ref{tbl-baseline} summarizes baseline demographic and clinical
characteristics by condition.

\begin{longtable}[]{@{}
  >{\raggedright\arraybackslash}p{(\linewidth - 6\tabcolsep) * \real{0.1667}}
  >{\centering\arraybackslash}p{(\linewidth - 6\tabcolsep) * \real{0.2778}}
  >{\centering\arraybackslash}p{(\linewidth - 6\tabcolsep) * \real{0.2778}}
  >{\centering\arraybackslash}p{(\linewidth - 6\tabcolsep) * \real{0.2778}}@{}}
\caption{Baseline demographic and clinical characteristics by
condition.}\label{tbl-baseline}\tabularnewline
\toprule\noalign{}
\begin{minipage}[b]{\linewidth}\raggedright
Characteristic
\end{minipage} & \begin{minipage}[b]{\linewidth}\centering
Treatment (\emph{n} = 121)
\end{minipage} & \begin{minipage}[b]{\linewidth}\centering
Control (\emph{n} = 133)
\end{minipage} & \begin{minipage}[b]{\linewidth}\centering
Total (\emph{N} = 254)
\end{minipage} \\
\midrule\noalign{}
\endfirsthead
\toprule\noalign{}
\begin{minipage}[b]{\linewidth}\raggedright
Characteristic
\end{minipage} & \begin{minipage}[b]{\linewidth}\centering
Treatment (\emph{n} = 121)
\end{minipage} & \begin{minipage}[b]{\linewidth}\centering
Control (\emph{n} = 133)
\end{minipage} & \begin{minipage}[b]{\linewidth}\centering
Total (\emph{N} = 254)
\end{minipage} \\
\midrule\noalign{}
\endhead
\bottomrule\noalign{}
\endlastfoot
Age, \emph{M} (\emph{SD}) & 35.3 (9.8) & 37.3 (11.5) & 36.4 (10.8) \\
Sex, \emph{n} (\%) & & & \\
~~Female & 72 (69.2\%) & 86 (69.9\%) & 158 (69.6\%) \\
~~Male & 31 (29.8\%) & 37 (30.1\%) & 68 (30.0\%) \\
~~Prefer not to say & 1 (1.0\%) & 0 (0.0\%) & 1 (0.4\%) \\
Ethnicity, \emph{n} (\%) & & & \\
~~White & 73 (70.2\%) & 88 (71.5\%) & 161 (70.9\%) \\
~~Black & 10 (9.6\%) & 20 (16.3\%) & 30 (13.2\%) \\
~~Asian & 8 (7.7\%) & 4 (3.3\%) & 12 (5.3\%) \\
~~Mixed & 11 (10.6\%) & 10 (8.1\%) & 21 (9.3\%) \\
~~Other & 2 (1.9\%) & 1 (0.8\%) & 3 (1.3\%) \\
Months since breakup, \emph{M} (\emph{SD}) & 15.0 (23.3) & 20.4 (33.7) &
17.8 (29.3) \\
Relationship length, \emph{n} (\%) & & & \\
~~Less than 6 months & 16 (13.2\%) & 13 (9.8\%) & 29 (11.4\%) \\
~~1 year & 25 (20.7\%) & 16 (12.0\%) & 41 (16.1\%) \\
~~2 years & 23 (19.0\%) & 22 (16.5\%) & 45 (17.7\%) \\
~~3 years & 15 (12.4\%) & 19 (14.3\%) & 34 (13.4\%) \\
~~More than 3 years & 42 (34.7\%) & 63 (47.4\%) & 105 (41.3\%) \\
Breakup initiator, \emph{n} (\%) & & & \\
~~Me & 36 (29.8\%) & 27 (20.3\%) & 63 (24.8\%) \\
~~A bit more me & 22 (18.2\%) & 22 (16.5\%) & 44 (17.3\%) \\
~~Mutual & 19 (15.7\%) & 28 (21.1\%) & 47 (18.5\%) \\
~~A bit more them & 21 (17.4\%) & 27 (20.3\%) & 48 (18.9\%) \\
~~Them & 23 (19.0\%) & 29 (21.8\%) & 52 (20.5\%) \\
Still in contact with ex, \emph{n} (\%) & 49 (40.5\%) & 58 (43.6\%) &
107 (42.1\%) \\
Currently in new relationship, \emph{n} (\%) & 40 (33.1\%) & 40 (30.1\%)
& 80 (31.5\%) \\
BDS total score, \emph{M} (\emph{SD}) & 35.3 (10.2) & 36.0 (11.5) & 35.6
(10.9) \\
ECRS Anxiety, \emph{M} (\emph{SD}) & 4.3 (1.2) & 4.4 (1.3) & 4.3
(1.2) \\
ECRS Avoidance, \emph{M} (\emph{SD}) & 3.1 (1.0) & 3.1 (1.0) & 3.1
(1.0) \\
\end{longtable}

\emph{Note.} All values in this table refer to the 254 baseline-valid
participants retained in the primary Python mixed-model sample. Age,
sex, and ethnicity were available from the merged Prolific exports for
227 participants because some participants had revoked data-sharing
consent on the platform; percentages for those rows therefore use
available data within each condition.

Demographic variables were available for
227 of the 254 primary-sample participants because some participants had
revoked data-sharing consent on Prolific. Most participants with
available demographic data were female and White, consistent with the
demographics of Prolific Academic's US and UK user base. Time since
breakup ranged from under one month to several years, and breakups had
been initiated by the participant, the partner, or both. About
two-fifths of participants were still in contact with their ex-partner,
and roughly one-third had entered a new relationship.

Groups did not differ at baseline on the primary outcome or the
preregistered attachment moderators. Baseline BDS scores were similar
across conditions (treatment: M = 35.29, SD = 10.23; control: M = 35.96,
SD = 11.53; \(p = .623\)). Attachment anxiety (treatment: M = 4.27;
control: M = 4.37; \(p = .518\)) and attachment avoidance (treatment: M
= 3.10; control: M = 3.06; \(p = .745\)) were also comparable. Time since breakup was numerically lower in the
treatment group (15.0 months) than in the control group (20.4 months),
but this difference did not reach significance (\(p = .134\)).

\subsection{User Experience}\label{user-experience-appendix}

User-experience ratings were collected after the day-0 protocol from
post-session-survey completers. Table~\ref{tbl-ux-perceptions} summarizes
perceptions of the system by condition. The mean recommendation rating
(0--10) was 7.92 (\(n = 94\), SD = 2.14) in the treatment group and 5.80
(\(n = 109\), SD = 2.42) in the control group, \(p < .001\),
\(d = 0.92\). Of the treatment participants, 61.7\% reported experiencing
a sudden insight or ``aha moment'' during the session, compared to
19.3\% of controls, \(p < .001\). At the 7-day follow-up, insight rates
remained higher in the treatment group among responders
(55.7\% vs.~22.8\%), \(p < .001\).

\begin{longtable}[]{@{}
  >{\raggedright\arraybackslash}p{(\linewidth - 4\tabcolsep) * \real{0.2308}}
  >{\centering\arraybackslash}p{(\linewidth - 4\tabcolsep) * \real{0.3846}}
  >{\centering\arraybackslash}p{(\linewidth - 4\tabcolsep) * \real{0.3846}}@{}}
\caption{User experience perceptions by
condition.}\label{tbl-ux-perceptions}\tabularnewline
\toprule\noalign{}
\begin{minipage}[b]{\linewidth}\raggedright
Item
\end{minipage} & \begin{minipage}[b]{\linewidth}\centering
Treatment (\emph{n} = 94), \emph{M} (\emph{SD})
\end{minipage} & \begin{minipage}[b]{\linewidth}\centering
Control (\emph{n} = 109), \emph{M} (\emph{SD})
\end{minipage} \\
\midrule\noalign{}
\endfirsthead
\toprule\noalign{}
\begin{minipage}[b]{\linewidth}\raggedright
Item
\end{minipage} & \begin{minipage}[b]{\linewidth}\centering
Treatment (\emph{n} = 94), \emph{M} (\emph{SD})
\end{minipage} & \begin{minipage}[b]{\linewidth}\centering
Control (\emph{n} = 109), \emph{M} (\emph{SD})
\end{minipage} \\
\midrule\noalign{}
\endhead
\bottomrule\noalign{}
\endlastfoot
Ease of use (UMUX-Lite) & 6.51 (0.88) & 6.37 (0.78) \\
Capabilities (UMUX-Lite) & 5.97 (1.29) & 5.28 (1.07) \\
Safety and appropriateness & 6.04 (1.15) & 5.80 (1.00) \\
Empathy and support & 6.26 (1.08) & 4.78 (1.42) \\
Trust and credibility & 5.89 (1.20) & 5.03 (1.22) \\
Anthropomorphism & 5.72 (1.33) & 4.56 (1.66) \\
\end{longtable}

UMUX-Lite ease-of-use scores were similarly high in both
groups (treatment: M = 6.51; control: M = 6.37; \(d = 0.17\)). The
capabilities item showed a wider gap (treatment: M = 5.97; control: M =
5.28; \(d = 0.58\)), with the chatbot conversation rated as meeting
participant needs more fully than the survey-only experience. Perceived
safety was high in both groups (treatment: M = 6.04; control: M = 5.80;
\(d = 0.23\)). Across the remaining perception items, treatment
participants consistently rated the chatbot higher than control
participants on empathy and support (\(d = 1.16\)), trust and
credibility (\(d = 0.71\)), and anthropomorphism (\(d = 0.77\)).

\emph{Note.} All items rated on a 7-point scale (1 = \emph{strongly
disagree}, 7 = \emph{strongly agree}). Treatment items referred to ``the
chatbot''; control items referred to ``the app.''

\end{document}